\documentclass[
 reprint,
 superscriptaddress,
 longbibliography,
 nostarredaffiliations,
 nofootinbib,
 amsmath,amssymb,
]{revtex4-2}
\usepackage[utf8]{inputenc} 
\usepackage[T1]{fontenc}
\usepackage{lmodern} 
\usepackage{graphicx}
\usepackage{dcolumn}
\usepackage{bm}
\usepackage[colorlinks=true,linkcolor=blue,urlcolor=blue,citecolor=blue]{hyperref}
\usepackage{multirow}

\usepackage{color}
\usepackage{xcolor}
\usepackage{times}
\usepackage{braket}
\usepackage{siunitx}
\usepackage{natbib} 
\usepackage{tabularx}
\usepackage{ulem}
\usepackage{newfloat}

\DeclareFloatingEnvironment[
fileext=lof, 
listname={List of Extended Data Figures}, 
name={Extended Data Fig}, 
placement=htp 
]{extfig}

\newcommand{\Rb}{\ensuremath{^{87}\text{Rb}~}}

\newcommand{\degree}{\ensuremath{^{\circ}}}
\newcommand{\wavr}[1]{\ensuremath{\braket{#1}_{\mathrm{w}}}}
\newcommand{\revise}[1]{{\color{red} #1}}

\begin{document}

\title{Realization of a doped quantum antiferromagnet in a Rydberg tweezer array}

\author{Mu~Qiao$^{\ddagger}$}
\altaffiliation{These authors contributed equally to this work.}
\affiliation{
Universit\'{e} Paris-Saclay, Institut d'Optique Graduate School, CNRS, Laboratoire Charles Fabry, 91127 Palaiseau Cedex, France
}
\email{mu.q.phys@gmail.com}
\author{Gabriel~Emperauger}
\altaffiliation{These authors contributed equally to this work.}
\affiliation{
Universit\'{e} Paris-Saclay, Institut d'Optique Graduate School, CNRS, Laboratoire Charles Fabry, 91127 Palaiseau Cedex, France
}
\author{Cheng~Chen}
\altaffiliation{These authors contributed equally to this work.}
\affiliation{
Universit\'{e} Paris-Saclay, Institut d'Optique Graduate School, CNRS, Laboratoire Charles Fabry, 91127 Palaiseau Cedex, France
}
\affiliation{
Institute of Physics, Chinese Academy of Sciences, Beijing 100190, China
}

\author{Lukas~Homeier}
\altaffiliation{These authors contributed equally to this work.}
\affiliation{JILA and Department of Physics, University of Colorado, Boulder, CO, 80309, USA}
\affiliation{Center for Theory of Quantum Matter, University of Colorado, Boulder, CO, 80309, USA}

\author{Simon~Hollerith}
\altaffiliation{These authors contributed equally to this work.}
\affiliation{
Department of Physics, Harvard University, Cambridge, Massachusetts 02138, USA
}

\author{Guillaume~Bornet}
\affiliation{
Universit\'{e} Paris-Saclay, Institut d'Optique Graduate School, CNRS, Laboratoire Charles Fabry, 91127 Palaiseau Cedex, France
}
\author{Romain~Martin}
\affiliation{
Universit\'{e} Paris-Saclay, Institut d'Optique Graduate School, CNRS, Laboratoire Charles Fabry, 91127 Palaiseau Cedex, France
}
\author{Bastien~G\'ely}
\affiliation{
Universit\'{e} Paris-Saclay, Institut d'Optique Graduate School, CNRS, Laboratoire Charles Fabry, 91127 Palaiseau Cedex, France
}
\author{Lukas~Klein}
\affiliation{
Universit\'{e} Paris-Saclay, Institut d'Optique Graduate School, CNRS, Laboratoire Charles Fabry, 91127 Palaiseau Cedex, France
}
\author{Daniel~Barredo}
\affiliation{
Universit\'{e} Paris-Saclay, Institut d'Optique Graduate School, CNRS, Laboratoire Charles Fabry, 91127 Palaiseau Cedex, France
}
\affiliation{
Nanomaterials and Nanotechnology Research Center (CINN-CSIC), Universidad de Oviedo (UO), Principado de Asturias, 33940 El Entrego, Spain
}
\author{Sebastian~Geier}
\affiliation{
Physikalisches Institut, Universit\"at Heidelberg, Im Neuenheimer Feld 226, 69120 Heidelberg, Germany
}
\author{Neng-Chun~Chiu}
\affiliation{
Department of Physics, Harvard University, Cambridge, Massachusetts 02138, USA
}
\author{Fabian~Grusdt}
\affiliation{
Department of Physics and Arnold Sommerfeld Center for Theoretical Physics (ASC), Ludwig-Maximilians-University at M\"{u}nchen, Theresienstr. 37, M\"{u}nchen D-80333, Germany
}
\affiliation{Munich Center for Quantum Science and Technology (MCQST), Schellingstr. 4, M\"unchen D-80799, Germany}
\author{Annabelle~Bohrdt}
\affiliation{University of Regensburg, Universit\"atsstr. 31, Regensburg D-93053, Germany}
\affiliation{Munich Center for Quantum Science and Technology (MCQST), Schellingstr. 4, M\"unchen D-80799, Germany}
\author{Thierry~Lahaye}
\affiliation{
Universit\'{e} Paris-Saclay, Institut d'Optique Graduate School, CNRS, Laboratoire Charles Fabry, 91127 Palaiseau Cedex, France
}

\author{Antoine~Browaeys}
\email{antoine.browaeys@institutoptique.fr}
\affiliation{
Universit\'{e} Paris-Saclay, Institut d'Optique Graduate School, CNRS, Laboratoire Charles Fabry, 91127 Palaiseau Cedex, France
}

\date{\today}
\begin{abstract}
Doping an antiferromagnetic Mott insulator is central to our understanding of a variety of phenomena in strongly-correlated electrons, including high-temperature superconductors~\cite{Lee2006,Bohrdt2021}. To describe the competition between tunneling~$t$ of hole dopants and antiferromagnetic (AFM) spin interactions~$J$, theoretical and numerical studies often focus on the paradigmatic $t$-$J$~model~\cite{Auerbach1994}, and the direct analog quantum simulation of this model in the relevant regime of high-particle density has long been sought~\cite{junye_tJ,Gorshkov_2011_tJVW}.
Here, we realize a doped quantum antiferromagnet with next-nearest neighbour (NNN) tunnelings~$t'$~\cite{qin2020absence,jiang_t'_2019,jiang_t'_2024,Xu2024,Bespalova2024} and hard-core bosonic holes~\cite{Bohrdt2024} using a Rydberg tweezer platform.
We utilize coherent dynamics between three Rydberg levels, encoding spins and holes~\cite{Lukas_2024_tJ}, to implement a tunable bosonic $t$-$J$-$V$~model allowing us to study previously inaccessible parameter regimes. We observe dynamical phase separation between hole and spin domains for $|t/J|\ll 1$, and demonstrate the formation of repulsively bound hole pairs in a variety of spin backgrounds. The interference between NNN tunnelings~$t'$ and perturbative pair tunneling gives rise to light and heavy pairs depending on the sign of $t$. Using the single-site control allows us to study the dynamics of a single hole in 2D square lattice (anti)ferromagnets.
The model we implement extends the toolbox of Rydberg tweezer experiments beyond spin-1/2 models~\cite{Browaeys2020_NP} to a larger class of $t$-$J$ and spin-$1$~models~\cite{Mogerle2024,Liu2024}.
\end{abstract}

\maketitle

The low-energy and out-of-equilibrium properties of materials with 
strong electronic interactions are notoriously difficult to model, 
both theoretically and numerically~\cite{Scalapino2012,Proust2019}. 
Although cuprate superconductors have been discovered almost four decades ago, 
a full understanding of the mechanisms underlying 
unconventional, high-temperature superconductivity is lacking~\cite{Bednorz1986,Keimer2015}. 
The most intricate regime in the paradigmatic Fermi-Hubbard or $t$-$J$~model~\cite{Auerbach1994} 
arises upon doping the antiferromagnetic (AFM) Mott insulator~\cite{Lee2006}, 
where the kinetic motion~$t$ of 
holes competes with the magnetic ordering of spins.
Understanding this interplay between magnetism and charge dynamics at low doping is 
important to develop phenomenological models of high-temperature superconductivity. 

The controlled quantum simulation of many-body systems~\cite{Bloch2008_review} 
has gained significant importance in the investigation of strongly-correlated materials, 
and ubiquitous phenomena, such as long-range AFM in the square~\cite{Mazurenko2017} and cubic lattice~\cite{Shao2024}, 
pairing in one-dimensional ladders~\cite{Hirthe2023} or Nagaoka ferromagnetism in the triangular 
lattice~\cite{Lebrat2024,Prichard2024} have been observed, among others~\cite{Bohrdt2021}.
So far, the leading platform for the quantum simulation of doped quantum magnets 
has been ultracold fermionic atoms in optical lattices~\cite{Duan2003,Gross2017}. In this system, the AFM spin coupling~$J=4t^2/U$ originates perturbatively from 
strong repulsive Hubbard interactions~$U\gg t$, thereby restricting the parameter space to~$t \gg J$ in the effective $t$-$J$~model.

\begin{figure*}
\mbox{}
\includegraphics[width=\textwidth]{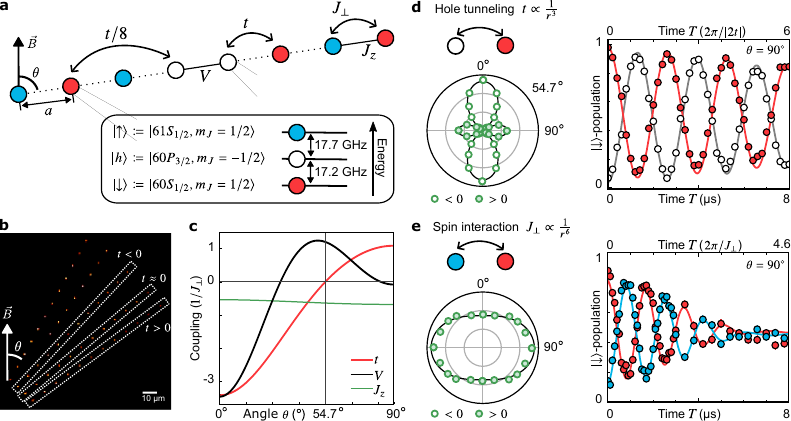}
\caption{\label{fig:setup} 
{\bf Implementation of $t$-$J$-$V$ model in a Rydberg tweezer array.}
\textbf{a,} The three Rydberg states $\ket{61S_{1/2}}$, $\ket{60S_{1/2}}$, and $\ket{60P_{3/2}}$ states encode spin up 
($\ket{\uparrow}$), spin down ($\ket{\downarrow}$), and hole ($\ket{h}$) respectively. 
The atomic pair interactions give rise to tunnelings ($t\propto r^{-3}$), and spin-spin ($J_{\perp}$, $J_z$) and hole-hole ($V$) interactions. 
\textbf{b,} Fluorescence image of one-dimensional atomic arrays oriented at different angles $\theta$ relative 
to the quantization axis defined by the magnetic field $\vec{B}$. For an experiment
corresponding to a given $\theta$, only the relevant chain is filled with atoms. 
\textbf{c,} Calculated angular dependence of the interaction strengths normalized 
to $J_{\perp}$, showing the tunability of the different terms in the $t$-$J$-$V$ Hamiltonian at distance $a=9.9\,\mu$m. \textbf{d,e}, Measured angular dependence of interaction strength and coherent evolution using isolated atom pairs. 
The hole tunneling~$t$ (d) exhibits strong angular variation with vanishing amplitude 
at the magic angle $\theta_m = 54.7\degree$, while the spin-spin interaction~$J_{\perp}$ (e) shows weaker angular modulation. 
Filled (empty) green circles in the left panels indicate positive (negative) interaction amplitudes, black lines represent theoretically calculated interaction values. 
Right panels show the $\ket{\downarrow}$-population of two atoms undergoing coherent state exchange, the color indicates each atom's initial state.
Solid lines show numerical simulations including thermal atomic motion \cite{GEinprep}. 
}
\end{figure*}

While this is the most relevant regime for 
strongly-correlated electrons, it prevents probing many of the other theoretically 
predicted phases such as phase separation~\cite{emery1990separation,Boninsegni2001},
or models with bosonic hole dopants~\cite{Boninsegni2001,Sun2021,Jepsen2021}, 
in which recent numerical studies have predicted stripe-ordered~\cite{Lukas_2024_tJ,Harris2024} and paired 
phases~\cite{Zhang2024} suggesting a common mechanism underlying AFM $t$-$J$~models.
Recently, the preparation of negative temperature states in a Bose-Hubbard model studied the dynamics of a single-hole in a bosonic AFM~\cite{Bohrdt2024}.
Notably, another promising route to directly realize $t$-$J$-$V$-$W$~models with 
dipolar AFM interactions, i.e. including more general hole-hole (spin-hole) interactions $V$ ($W$), has recently been achieved in itinerant, fermionic polar molecules in optical 
lattices~\cite{junye_tJ,Gorshkov_2011_tJVW} enabling access to a greater parameter space. The implementation of tunnelings~$t$ beyond nearest-neighbour (NN) sites in optical lattices remains challenging but is desired to test predictions, such as their importance for superconductivity,
which has received significant attention recently~\cite{qin2020absence,jiang_t'_2019,jiang_t'_2024,Xu2024,Bespalova2024}. To our knowledge, no direct realization of the $t$-$J$~model has been achieved in the high particle filling regime with single-site resolution, nor has tunneling beyond nearest neighbors been implemented.

As recently proposed in Ref.~\cite{Lukas_2024_tJ}, we encode the hole and spins into three 
Rydberg states of \Rb atoms. The dipole-dipole and the van der Waals (vdW) 
interactions between atoms implement a hard-core bosonic $t$-$J$-$V$ model, with at 
most one particle per site. The $1/r^3$ dipole-dipole interactions yield hole 
tunneling $\propto t$, and the $1/r^6$ vdW interactions realize 
magnetic spin coupling $\propto J_{\perp}, J_{z}$, and hole-hole interactions $\propto V$.  
In a 1D chain, the tweezers setup allows us to tune the ratio between hole tunneling $t$, 
spin interaction $J_{\perp,z}$, and hole-hole interaction $V$ over a large range by changing the orientation of the chain
with respect to the quantization axis. In a 2D array, the ratio $t/J$ can be tuned by varying the site distance. In both cases, 
we can access the regimes $t\ll J_{\perp}$ and $t\gg J_{\perp}$. 
In our implementation, we study
the dynamical properties of doped 1D and 2D quantum magnets with long-range dipolar tunneling.

In this  model, the interplay between tunneling~$t$ and AFM spin flip-flop interactions~$J_
\perp$ introduces a ``sign problem'', even when the exchange statistics of holes is bosonic~\cite{Zhang2024}. At high particle density, this frustration leads to strong spin-charge correlations~\cite{Lukas_2024_tJ}, such as paired phases~\cite{Zhang2024}. While the statistics plays a crucial role for collective properties, our model explores underlying mechanisms, including magnetically-mediated pairing mechanisms~\cite{Siller2001,OMahony2022} or kinetic magnetism~\cite{Morera2024}. Furthermore, the broad tunability of the parameters opens the door a wide range of exotic phenomena~\cite{Sous2020}.

Our results are fourfold. First, in a 1D chain, using our ability to prepare initial product states and to tune the ratio of tunneling strength $t$ to magnetic spin interaction $J$, we observe a dynamical phase separation between holes and spins and probe the properties of repulsively bound hole pairs. Second, we probe the interplay between NNN tunneling~$t'$ and the perturbative pair tunneling induced by NN tunneling~$t$. Depending on the sign of the hole tunneling $t'$, this leads to constructive or destructive interference, which allows us to control the effective mass of the hole pairs, as well as their mobility. 
Third, we investigate the influence of the spin background on the pair's mass and binding energy. Last, we perform single-hole experiments in a 2D array both for an FM and AFM spin background. In particular, in the FM case, we observe the influence of the dipolar tail of the interactions. 

Our experimental setup relies on 1D and 2D arrays of individual \Rb atoms held in optical tweezers with an intersite spacing~$a$. 
As shown in Fig.\,\ref{fig:setup}a and b, 
we encode the effective spin-1/2 and the hole in three Rydberg states: 
$\ket{\downarrow}=\ket{60S_{1/2},m_{J}=1/2}$, $\ket{\uparrow}=\ket{61S_{1/2},m_{J}=1/2}$, 
and $\ket{h}=\ket{60P_{3/2},m_{J}=-1/2}$. 
The resonant interaction between states of different parity implements dipolar tunneling with amplitude $t_{\sigma} \propto (1-3\cos^2\theta)$~\cite{Barredo2015}, 
where $\theta$ is the angle between the interatomic vector and the quantization axis 
defined by a 46\,G magnetic field, either in the plane of the array (1D case, see Fig.\,\ref{fig:setup}b and e) or perpendicular to it (2D case), 
and $\sigma\in\{\uparrow,\downarrow\}$ highlights a slight spin dependence of the hole tunneling ($t_{\uparrow}/t_{\downarrow}\approx0.95$). In the following, we set $t= (t_{\uparrow}+t_{\downarrow})/2$. 
The spin-spin interactions arise from the off-diagonal vdW interactions 
$\propto r_{ij}^{-6}$ between the pair states $\ket{60S, 61S}$~\cite{GEinprep}, 
giving rise to AFM XY-type couplings $J_\perp$; 
in addition, the diagonal vdW interactions between all pairs of Rydberg states 
leads to Ising spin-spin $J_{z}$ and hole-hole $V$ interactions. 
The calculated angular dependencies of all the interactions are non-universal and state-dependent, see Fig.\,\ref{fig:setup}c \cite{Wadenpfuhl2024}.
The atomic pair interaction can be expressed in the language of a bosonic $t$-$J$-$V$ model given by
\cite{Lukas_2024_tJ, Gorshkov_2011_tJVW}:
\begin{equation}
\label{_Hamiltonian}
\begin{aligned}
&\hat{H}_{tJV} = \hat{H}_{t}+\hat{H}_{J}+\hat{H}_{V}\ ,  \\
&\hat{H}_{t} =-\sum_{i<j}\sum_{\sigma=\downarrow,\uparrow
}\frac{t_{\sigma}}{{r}_{ij}^3}\left(\hat{a}_{i, \sigma}^{\dagger} \hat{a}_{j, h}^{\dagger} \hat{a}_{i, h} \hat{a}_{j, \sigma}+\text { h.c. }\right), \\
&\hat{H}_{J} = \sum_{i<j}\frac{1}{{r}_{ij}^6}\left[J^{z}\hat{S}_i^z \hat{S}_j^z+\frac{J_{\perp}}{2}\left(\hat{S}_i^{+} \hat{S}_j^{-}+\text {h.c. }\right)\right]\ , \\
&\hat{H}_{V} =\sum_{i<j}\frac{V}{{r}_{ij}^6} \hat{n}_i^h \hat{n}_j^h\ .
\end{aligned}
\end{equation}
where we have set $\hbar=1$ and expressed distances $r_{ij}$ in units of the lattice spacing~$a$. The exact mapping contains additional boundary terms and spin-hole interactions~$W$ that we find to be numerically negligible (see Methods Extended Data Table \ref{table:bosonic-tJ-Ryd-},\, \ref{table:tJVW_interactions}). Although our observations can be qualitatively described by Eq.~(\ref{_Hamiltonian}), we include all terms in our numerical simulations. $\hat{H}_{tJV}$ consists of two parts: $\hat{H}_{t}+\hat{H}_{J}$ describes a hard-core bosonic $t$-$J$ model, and $\hat{H}_{V}$ represents the hole-hole interaction with interaction strengths~$t,J_\perp,J_z$ and~$V$ shown in Fig.\,\ref{fig:setup}c. 
The $t$-$J$ component comprises a tunneling term for particles and a magnetic XXZ interaction. 
The operators $\hat{a}_{j,\sigma}^{\dagger}$ and $\hat{a}_{j,h}^{\dagger}$ represent Schwinger bosons that
create a spin~$\sigma$ and a hole at site~$j$, respectively. 
These operators obey bosonic commutation relations for different sites, 
with an additional hard-core constraint $\sum_\sigma \hat{a}_{j,\sigma}^{\dagger}\hat{a}_{j,\sigma} + \hat{a}_{j,h}^{\dagger}\hat{a}_{j,h} = 1$ on each site. 
We denote the hole density 
operator at site $j$ by $\hat n_j^h=\hat{a}_{j,h}^{\dagger}\hat{a}_{j,h}$.
The spin-1/2 operators $\hat{S}_{j}^{z}$ and $\hat{S}_{j}^{\pm} = \hat{S}_{j}^{x} \pm i \hat{S}_{j}^{y}$ at site~$j$ only act on the states $\ket{\uparrow}$ and $\ket{\downarrow}$ with $\hat{S}_{j}^{\gamma}=\frac{1}{2}\hat{a}^\dagger_{j,\alpha}\sigma^{\gamma}_{\alpha\beta}\hat{a}_{j,\beta}$ and Pauli matrices $\sigma^\gamma$ ($\gamma=x,y,z$). 
The Hamiltonian (\ref{_Hamiltonian}) features a $U(1)$ conservation of hole dopants 
$\hat{N}_{h}=\sum_{j}\hat{n}_{j}^{h}$, and conserves the total magnetization 
$\hat{S}_{\rm{tot}}^{z}=\sum_{j}\hat{S}_{j}^{z}$. 

\begin{figure*}
    \includegraphics[width=\textwidth]{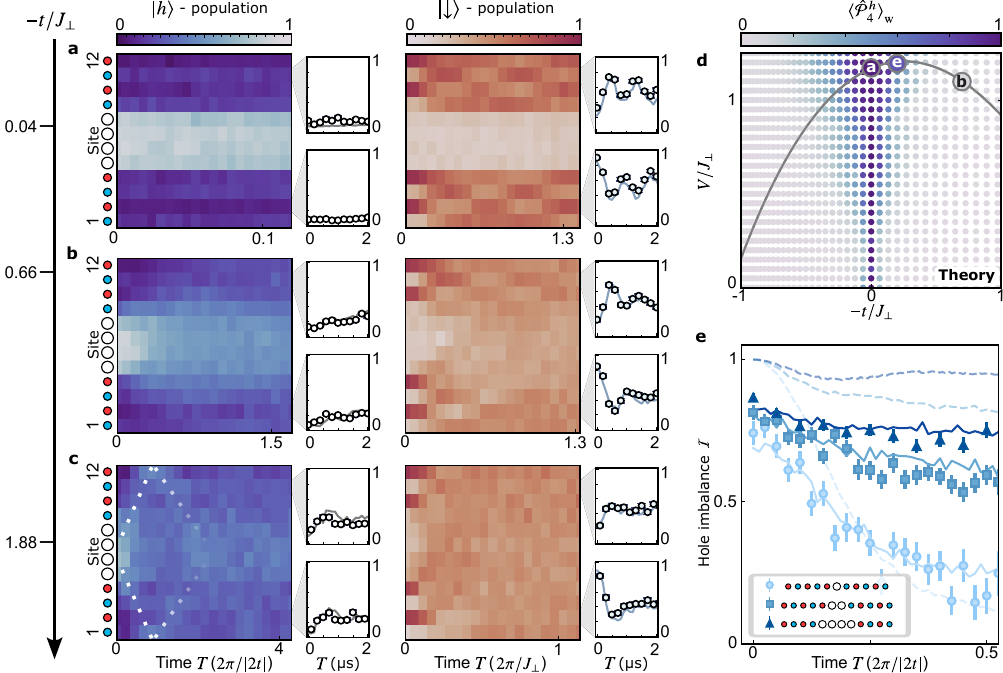}
    \caption{\textbf{Dynamical phase separation in the bosonic $t$-$J$-$V$ model.}
    \textbf{a-c}, Time evolution of site-resolved $\ket{h}$ (hole) and 
    $\ket{\downarrow}$ (spin-down) populations for varying ratios $t/J_\perp$: 
    \textbf{a}, $t/J_\perp \approx -0.043(5)$ (measured); 
    \textbf{b}, $t/J_\perp \approx -0.66$ (calculated); 
    \textbf{c}, $t/J_\perp \approx -1.88$ (calculated). 
    Initial states are indicated by 
    colored dots in the leftmost column. Insets show population dynamics for 
    selected sites (2nd and 11th), comparing experimental data 
    (points, error bars denote one standard error) with numerical simulations including experiment imperfections (solid lines). 
    \textbf{d}, Weighted average of NN 4-hole correlations $\wavr{\hat{\mathcal{P}}^{h}_{4}}$
    for the initial state 
    $\ket{\psi_{0}}=\ket{\uparrow\downarrow\uparrow\downarrow hhhh\uparrow\downarrow\uparrow\downarrow}$ (see text)
    as a function of $t/J_{\perp}$ and $V/J_{\perp}$, calculated from theory. A value of $\langle\hat{\mathcal{P}}_4^h\rangle=1$ indicates binding of all holes into one cluster. The gray curve shows experimentally accessible parameters through variation of angle $\theta$. 
    \textbf{e}, Evolution of the hole imbalance $\cal I$ for various doping densities 
    (1 hole, 2 holes, 4 holes, see inset) at $t/J_\perp \approx -0.19$ and $V/J_\perp \approx 1.2$ (corresponding to an angle $\theta=51.7^\circ$). 
    We post-selected experimental data containing hole numbers of 1, 2 and 4 respectively. Solid (dashed) lines are comparisons to numerical simulations including (without) experimental imperfections.
    }
    \label{fig:dynamics}
\end{figure*}

In this Rydberg encoding scheme, the tunneling term exhibits a $1/r^3$ dipolar behavior, 
whereas the magnetic spin interaction decays as $1/r^6$. 
The different power-law scaling of the tunneling and spin interactions with distance combined with the angular
dependence of $t$ allows tuning the ratio $t/J_{\perp}$ over a wide range, see Fig.\,\ref{fig:setup}c. 
At the magic angle $\theta_{ij}\approx 54.7^{\circ}$, 
the tunneling vanishes ($t=0$), placing the system in the $|t|\ll J_\perp$ regime; 
increasing the interatomic distance $r_{ij}$ leads to $|t|\gg J_\perp$ due to the 
different spatial decay of these interactions.
In our experiment, for $a=9.9\,\mu$m corresponding to the 1D chain studied below, antiferromagnetic XY interactions with numerically calculated strength $J_{\perp} = 2\pi \times 692~\text{kHz}$ coexist with 
ferromagnetic Ising interactions $J_{z} = -2\pi \times 443~\text{kHz}$ and hole-hole interaction $V=2\pi\times819~\text{kHz}$ at $\theta = 54.7^{\circ}$. The hole tunneling amplitude can 
be tuned between $|t|=2\pi\times (0 \dots 1)$\,MHz.
We have confirmed experimentally the calculated values and angular dependencies of $J_\perp$
and $t_{\downarrow}$ using pairs of atoms
prepared either in a $\ket{\downarrow h}$ or $\ket{\downarrow\uparrow}$ configuration
and observing the corresponding resonant exchange dynamics between the two atoms. 
The results are summarized in 
Fig.\,\ref{fig:setup}d and e. We discuss the experimental imperfections in the Methods and in Ref.~\cite{GEinprep}.  
For all the many-body experiments reported below, the starting point 
is a product state, prepared by site-dependent light shifts and microwave rotations (see Methods). 

In a first set of experiments, we demonstrate the tunability of the platform 
by investigating, in a 1D chain, the interplay 
between spin dynamics and hole propagation in various regimes. 
To do so, we vary the angle $\theta$ of the tweezer chain with respect to the in-plane magnetic field, thus tuning the ratio of 
tunneling amplitude $t$ to spin-spin interaction $J_\perp$. 
We initialize a 12-atom chain in a N\'eel antiferromagnetic configuration 
along $z$ with four holes positioned at its center: $\ket{\psi_{0}}=\ket{\uparrow\downarrow\uparrow\downarrow hhhh\uparrow\downarrow\uparrow\downarrow}$, 
see Fig.\,\ref{fig:dynamics}. For this initial state with a high energy density, we expect dynamics of both, holes and spins. 

We then suddenly turn off the light shift, 
let the system evolve freely under $\hat{H}_{tJV}$, 
and measure the state's evolution as a function of time. 
Fig.\,\ref{fig:dynamics}a-c present the time evolution of the hole density 
$\braket{\hat{n}^{h}_{i}}$ and the spin density $\braket{\hat{n}^{\downarrow}_{i}}$ for three angles. Close to the magic angle $\theta = 54.7^\circ$ (Fig.\,\ref{fig:dynamics}a), 
where $|t/J_{\perp}| \ll 1$ (experimentally, $|t|= 2\pi\times30(10)$~kHz),
we observe a separation of hole and spin domains. 
The holes, initially prepared in the center, 
form a static domain that remains separated from the surrounding spins. 
This situation exhibits minimal 
hole diffusion while still featuring coherent spin oscillations. 
We compare the results to numerical simulations including experimental imperfections (see Methods), 
finding good agreement between the two. 
Using this simulation, we extrapolate  
that the separation of spin and charge domains survives up to times~$T$ longer than the one achieved experimentally.
When we decrease $\theta$ to $45^\circ$ ($t/J_{\perp} \approx -0.66$, $V/J_{\perp}\approx 1.1$ Fig.\,\ref{fig:dynamics}b) we still measure substantial occupation of the holes on their initial sites, although the domain wall begins to destabilize. Concurrently, we observe damped spin oscillations in the region where holes partially penetrate. Further reducing $\theta$ to $30^\circ$ ($t/J_{\perp} \approx -1.88$, $V/J_{\perp}\approx-0.18$) dramatically alters the holes' behavior as seen in Fig.\,\ref{fig:dynamics}c. In this strong tunneling regime, holes rapidly delocalize across the entire chain, and are reflected at the boundary as highlighted by the dotted lines in Fig.\,\ref{fig:dynamics}c (guide to the eye). The spin dynamics become erratic, indicating that magnetic order is suppressed by the holes' motion.

\begin{figure*}
    \mbox{}
    \includegraphics[width=\textwidth]{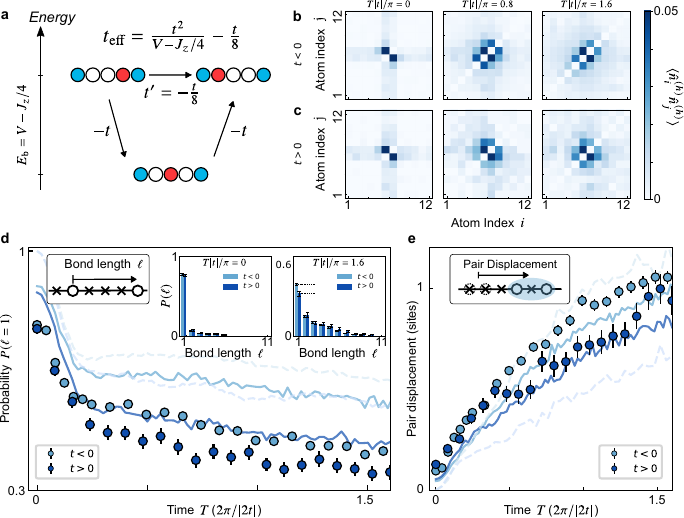}
    \caption{\label{fig:perturbative_tunneling} 
    {\bf Influence of the next-nearest neighbor tunneling on the dynamics of hole pairs.}
    \textbf{a}, Energy diagram illustrating the perturbative tunneling of bound pairs in a frozen spin background~$J_\perp = 0$, 
    where NN tunneling contributes $t^2/(V-J_z/4)$ (sign-independent) 
    and NNN tunneling contributes $t'=-t/8$ (sign-dependent). 
    \textbf{b,c}, hole-hole correlations $\langle \hat n^h_i\hat n_j^h\rangle $ at different times for ({\bf b}) $\theta=49.7^\circ$ ($t<0 $), and ({\bf c}) $\theta=59.7^\circ$ ($t>0$). 
    \textbf{d}, Time evolution of the probability of finding a bound state with bond length $\ell=1$ for positive 
    ($t>0$, dark blue) and negative ($t<0$, light blue) tunneling. Insets show the distribution of hole-hole separations at $T=0$ 
    and $T=4\,\mu$s.
    \textbf{e}, Time evolution of the pair displacement (see inset) for both $t>0$ and $t<0$. We restrict the analysis to configurations where the holes are separated by at most two bonds~$\ell \leq 2$.
    Dots are experimental data, solid lines represent numerical simulations with experimental error and dashed lines are the simulations in the absence of imperfections.
    Error bars denote one standard error of the mean.
    All the data of Fig.\,\ref{fig:perturbative_tunneling} are postselected, retaining only experiments containing two holes.
    }
\end{figure*}

To understand the origin of the initial phase separation between hole-rich and spin-rich regions, we need to consider the interplay between the processes described by $t$, $J_\perp$, $J_z$ and $V$.
In the limit of small tunneling $|t/J_z| \ll 1$ and $J_\perp =V =0$, we expect a trivial binding owing to the energy cost~$|J_z/4|$ of breaking a spin bond (we call the bond-breaking cost the binding energy). In the presence of spin fluctuations~$J_\perp$ but no hole-hole interaction ($V=0$), the hole cluster can be destabilized when the binding energy is absorbed by the spin background, especially for large system sizes. 
The observation of a stable phase separation points towards the stabilizing role of the hole-hole interaction $V$. To analyze this, we calculate the eigenstates~$\{\ket{n}\}$ of the chain, Eq.~\eqref{eq:spin-Hamiltonian_diag} (see Methods), for various values of 
$V/J_\perp$, and introduce a projector $\hat{\mathcal{P}}_4^h = \sum_j\hat{n}_j^h \hat{n}_{j+1}^h\hat{n}_{j+2}^h\hat{n}_{j+3}^h$ that extracts the hole-rich contribution of each eigenstate independent of the spin background. The weighted overlap~$\wavr{\hat{\mathcal{P}}^{h}_{4}}=\sum_{n}|\braket{n|\psi_{0}}|^2\braket{n|\hat{\mathcal{P}}_{4}^{h}|n}$ with our initial state~$\ket{\psi_{0}}$ is directly related to the diagonal ensemble, i.e. the expectation value~$\langle \hat{\mathcal{P}}_4^h \rangle$ after equilibration. 
The results are shown in Fig.\,\ref{fig:dynamics}d, 
where we observe that the range of $t/J_\perp$ for which we expect the dynamical phase separation increases with hole-hole repulsion.
This can be further understood by the effective mass of a cluster of holes.
For instance, a repulsively bound pair~\cite{Winkler_2006_repulsive} of adjacent holes~$\ket{..\!\downarrow\uparrow hh\downarrow\uparrow\!..}$ can only propagate through a second-order process where the pair temporarily breaks apart: one hole tunnels (amplitude $t$), followed by the breaking of a nearby $\uparrow,\downarrow$ pair (energy cost $V-J_z/4$), and subsequent recombination resulting in an effective tunneling amplitude $t^2/(V-J_z/4)$. Extending to a four-hole bound state, the hole cluster moves by a fourth-order perturbative process~$\propto t^4/(V-J_z/4)^3$. Therefore as system size increases, while keeping the density of holes fixed, we expect an increasingly heavy cluster of holes with a slower propagation. 


To confirm this experimentally,
we study the dynamics of 1-, 2- and 4-hole initial states
at fixed $t/J_{\perp}\approx-0.19$ and $V/J_\perp\approx 1.2$. 
To this end, we quantify the spatial separation between hole-rich and hole-free regions by the hole imbalance:
\begin{equation}
\mathcal{I}=
\frac{1}{N_{\text{hole}}}\sum_{i\in\text{A}}\braket{\hat{n}_{i}^{h}}
-\frac{1}{L-N_{\text{hole}}}\sum_{j\not\in\text{A}}\braket{\hat{n}_{j}^{h}}\ ,
\end{equation} 
where $N_{\text{hole}}$ is the number of holes, $L$ is the number of sites, and $\text{A}$ denotes the sites initially occupied by holes. Fig.\,\ref{fig:dynamics}e compares the time evolution of $\cal I$ for systems with $N_{\text{hole}}=1,2,\text{ and }4$. The data reveal a slowdown in the melting of the boundary between hole-rich and hole-free regions as the number of holes increases from the single hole case where the dynamics is purely governed by the tunneling amplitude~$t$, in agreement with our numerical simulations.

This study of phase separation exemplifies the tunability of the platform, enabling the investigation of previously inaccessible parameter regimes and initial states~\cite{Staszewski2024}. Our findings constitute an important step towards understanding phase separation that is found to compete with stripe order and the pseudogap in cuprate superconductors~\cite{white2000separation}, and is relevant to a broader class of strongly-correlated electron systems, such as magnetic oxides~\cite{Kagan2021}. Early numerical studies in hard-core bosonic $t$-$J$~models have further revealed phase separation~\cite{Boninsegni2001}, pointing towards a common underlying mechanism in doped antiferromagnets. Our comparison of 2- and 4-hole clusters shows that in the limit~$J, V \gg t$, the pairing of holes may occurs as a precursor to phase separation.

The perturbative argument presented above for the mobility of a 2-hole bound state neglects the fact that the tunneling amplitude $t$ results from 
a dipolar interaction allowing for direct, 
NNN tunneling~$t'$ of one of the holes constituting the pair across the other hole, as 
represented in Fig.\,\ref{fig:perturbative_tunneling}a. The propagation
of the bound pair from one site thus results from the interference between the second-order
coupling described above, and the direct tunneling with amplitude $t'=t/8$. 
This leads to an effective tunneling amplitude of the bound pair 
$t_{\rm eff} = \chi t^2/(V-J_{z}/4)-t/8$, where $\chi=\chi(T)$ is a time-dependent prefactor describing the spin fluctuations (see Methods). 
Depending on the sign of $t$, the interference can be destructive or constructive, allowing us to tune the pair's mass~$m_{\rm eff} \propto 1/(2t_{\rm eff})$ by varying the angle~$\theta$. To lowest order in~$t$ the binding energy~$E_b=V-J_{z}/4$ in our perturbative description is unaffected by the interference with NNN tunnelings~$t'$.

To reveal the effect of the tunneling's dipolar tail in the propagation 
of a bound pair, we analyze the dynamics for two different signs of $t$, controlled by the 
angle $\theta$. We choose two values of $\theta=49.7^{\circ}, 59.7^{\circ}$, corresponding 
respectively to ($t/J_\perp = -0.32$, $V/J_{\perp}=1.2$) and ($t/J_\perp = 0.29$, $V/J_{\perp}=1.0$). We plot in 
Fig.\,\ref{fig:perturbative_tunneling}b ($t>0$) and c ($t<0$) the non-connected 
hole-hole correlations $\langle \hat n^h_i\hat n_j^h\rangle $ following the preparation of a pair $hh$ at the 
center of the chain, after an evolution time $T=2\,\mu$s ($T=0.8\times\frac{2\pi}{|2t|}$) and $T=4\,\mu$s ($T=1.6\times\frac{2\pi}{|2t|}$). 
From the correlation maps, we observe that the pair remains essentially bound during the time evolution, since correlations propagate mainly along the diagonal. We also see that for $t<0$ ($\theta=49.7^{\circ}$) the bound pair spreads along the first diagonal of the correlation map (thin, long ellipse), while for $t>0$ ($\theta=59.7^{\circ}$) the correlations spread slower and also along the second diagonals (fat, short ellipse), suggesting different binding strength and delocalization speeds of the hole pairs.

\begin{figure}
    \mbox{}
    \includegraphics[width=0.47\textwidth]{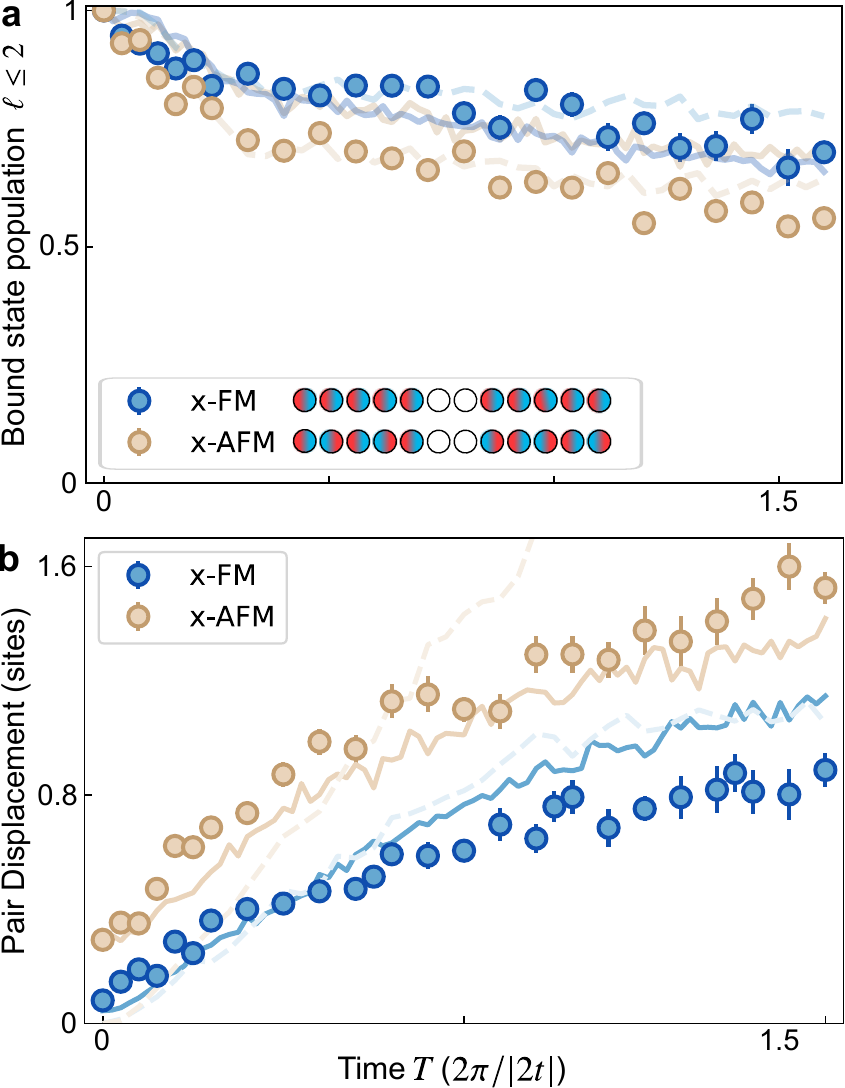}
    \caption{\label{fig:spin_background} {\bf Hole dynamics in different magnetic spin backgrounds.}
    \textbf{a}, Time evolution of the probability of finding two holes within two bonds of each other ($\ell \leq 2$) for ferromagnetic (x-FM, blue) and antiferromagnetic (x-AFM, yellow) spin backgrounds, normalized to the value at time $T=0$ to compensate for state preparation errors. 
    \textbf{b}, Time evolution of the pair displacement defined as in the caption of Fig.\,\ref{fig:perturbative_tunneling}{\bf d}.  Solid (dashed) lines show numerical simulations including (without) experimental imperfections in \textbf{a} and \textbf{b}. Error bars denote one standard error of the mean.
    }
\end{figure}

To quantify these observations, we first plot in the inset of Fig.\,\ref{fig:perturbative_tunneling}d the histogram of the distance between the two holes (bond length), at the initial time $T=0$ and after a time $T=1.6\times\frac{2\pi}{|2t|}$. As the probability of the bond length~$\ell=1$ at $T=1.6\times\frac{2\pi}{|2t|}$ is higher for $t<0$ than $t>0$, the binding is slightly tighter for the first case since $V$ is coincidentally stronger for $t<0$. This is consistent with the fact that the calculated binding energy $E_b=V+|J_{z}|/4$ (See Methods) is lower in the first case: $E_b = 2\pi \times 0.91$~MHz for $t<0$ and $E_b = 2\pi \times 0.85$~MHz for $t>0$. The Fig.\,\ref{fig:perturbative_tunneling}d shows the bond length~$\ell=1$ probability remains larger for the tightly bound pair than for the weakly bound pair throughout experimentally accessible times. We find that the numerical simulations using single-site state preparation errors qualitatively but not quantitatively describe the experimental data, possibly due to correlated errors in the initial state not included in the simulations (see Methods).

\begin{figure*}
    \mbox{}
    \includegraphics[width=\textwidth]{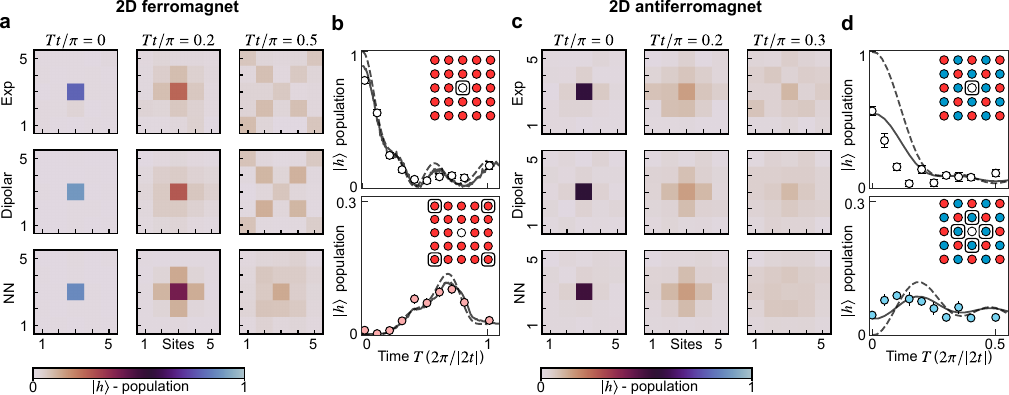}
    \caption{\label{fig:2d} {\bf Hole dynamics in 2D quantum magnets.}
    \textbf{a},~Spatial maps of the hole probability at various evolution times $T$, for a hole initialized in the 2D FM spin background $\ket{\downarrow}$. The top row shows the experimental data, the middle row is a numerical simulation with the dipolar couplings of the Rydberg interactions, and the bottom row considers the theoretical case of nearest-neighbor interactions.
    \textbf{b},~Time evolution of the hole probability at the initial center site (top row) and at a corner site (bottom row). Solid lines (resp. dashed lines) are simulations with (resp. without) experimental imperfections.
    \textbf{c,d},~Same as~(\textbf{a,b}), but in the case of an AFM spin background along~$z$. The bottom panel in~\textbf{d} now shows the hole population in a nearest-neighbor site to the initial position.
    }
\end{figure*}

Second, we characterize the pair mobility by measuring their center-of-mass displacement, defined as the distance between the final and initial center-of-mass positions. Here, we define the pairs as holes with bond length $\ell \leq 2$, to distinguish the bound state from the single particle background of the bimodal distribution visible in the histograms in Fig.\,\ref{fig:perturbative_tunneling}d at $T=1.6\times\frac{2\pi}{|2t|}$. The results are shown in Fig.\,\ref{fig:perturbative_tunneling}e and feature the expected asymmetry. They are also in qualitative (quantitative) agreement with numerical simulations without (with) errors. From Fig.\,\ref{fig:perturbative_tunneling}d and e, we conclude that pairs are lighter (heavier) and less (more) extended for $t<0$ ($t>0$). This behavior is a direct consequence of the $1/r^3$ tail in the tunneling amplitude of the holes, and would be absent for NN interactions. In particular, for NN interactions, the hole pairs would have the same mass~$m_{\rm eff} \propto 1/(2t_{\rm eff})$ according to the perturbative argument above.

In a last experiment in 1D, we investigate how different magnetic backgrounds influence the dynamics of the hole pairs. 
Specifically, we address two questions: does the spin background 
modify whether the pair remains bound, and if yes, how do the holes propagate? From the perturbative description of the bound pair, we infer that the binding energy and effective mass can be tuned by changing the spin contribution to the energy of the initial state. In particular, the binding energy~$E_b = V \pm J_\alpha/4$ depends on the relative orientation of spins ($\pm$) and direction ($J_\alpha = J_z,J_\perp$).
To probe this effect and investigate the influence of~$J_\perp$, we use a chain at $\theta=49.7^{\circ}$ and prepare two distinct spin configurations, now along $x$: 
an antiferromagnetic Néel order ($x$-AFM, $\ket{\leftarrow\rightarrow\leftarrow\rightarrow\leftarrow hh\rightarrow\leftarrow\rightarrow\leftarrow\rightarrow}$) 
and a ferromagnetic state ($x$-FM, $\ket{\leftarrow\leftarrow\leftarrow\leftarrow\leftarrow hh\leftarrow\leftarrow\leftarrow\leftarrow\leftarrow}$).
We again initialize two holes at the center and track the time evolution of their position. We first investigate the binding of holes by analyzing 
their separation distance, defining bound pairs as those separated by at most two bonds~$\ell \leq 2$. 
The holes remain bound in both magnetic backgrounds due to hole-hole interaction $V$, 
though slightly weaker in the $x$-AFM case, as shown in Fig.\,\ref{fig:spin_background}a. 
We find better agreement with ideal simulations, again indicating that our error model does not capture all experimental imperfections.

Next, we characterize the pair mobility by measuring the pair displacement. 
The results, shown in Fig.\,\ref{fig:spin_background}b, reveal that bound pairs propagate significantly faster in the $x$-AFM background 
compared to the $x$-FM case, in agreement with our numerical simulations. 
This behavior is consistent with the effective pair 
tunneling amplitude $\propto 1/(V\mp J_{\perp}/4)$ for $x$-AFM ($-$) and $x$-FM ($+$) backgrounds, respectively. 
Since $V$ and $J_{\perp}$ have the same sign, 
the pairs experience weaker binding in the $x$-AFM case. 
In contrast to the tightly bound, light pair and weakly bound, heavy pair revealed
in Fig.\,\ref{fig:perturbative_tunneling}, 
here we observe a tightly bound, heavy pair and a weakly bound, light pair.

Finally, we extend our study to investigate the dynamics of a single hole in two dimensions (2D). Our motivation is twofold. First, we aim at seeing the influence of the dipolar tail of the tunneling on the dynamics, in the case of a FM background on a square lattice. Second, we demonstrate the platform's ability to explore the hole dynamics in an AFM background, relevant for the understanding of doped AFM Mott insulators~\cite{Lee2006,Ji2021}.
In 2D, while the angular dependence of all couplings is fixed by setting the magnetic field
perpendicular to the atomic plane~$\theta=90\degree$ -- fixing the ratio $V/J_\perp=-0.07$, $J_z/J_\perp=-0.68$ -- we can tune the ratio $t/J_{\perp,z}$ by varying the lattice spacing, 
exploiting the different power-law of tunneling ($\propto r^{-3}$) and spin exchange ($\propto r^{-6}$) interactions. Here, we implement a $5\times5$ square array with spacing $a=12\,\mu$m  and $t/J_{\perp}=2$ ($t=2\pi \times 509 \,\mathrm{kHz}$) and initialize it in a polarized FM or AFM product state along $z$, with a single hole at the center. 

Fig.\,\ref{fig:2d}a shows the local hole occupation number in the FM case at two different times. There the dynamics reduces to that of a single particle (here the hole) tunneling in a 2D lattice with dipolar~$\propto t/r^3$ tunneling rate. The observed coherent evolution of the hole shows a distinctive interference pattern which is a fingerprint of the associated band structure:
in particular the hole occupation exhibits pronounced peaks at $T=0.5\times\frac{2\pi}{|2t|}$ ($T=0.515\,\mu$s) along the diagonal of the 2D array, a feature absent in the simulation including only NN interactions.
Fig.\,\ref{fig:2d}b focuses on the evolution of the hole occupation at the center and corner of the array, and compares simulations including experimental errors (solid) to the ideal simulations (dashed) with dipolar tails. 

The AFM case is presented in Fig~\ref{fig:2d}c. There, we expect that the hole's paths do not interfere because of their different spin backgrounds that keep memory of the trajectory~\cite{Ji2021}. Furthermore, the initial AFM state allows for strong fluctuations in both the charge (i.e. the hole) and spin sector due to tunneling~$t$ and spin flip-flop processes~$J_\perp$. 
In Fig.\,\ref{fig:2d}d we compare the evolution of the hole occupation with time-dependent matrix product state simulations to ideal (dashed) and error (solid) simulations incorporating measured state-preparation infidelities (see Methods). The error simulations (solid) are in good agreement with our experiment, by comparing the peak at the center and its adjacent sites at early evolution times; at later times both the numerical simulations and the experiment become featureless due to the high-energy initial state prepared in the experiment. The antiferromagnetic spin background makes different path distinguishable, consequently, interference between these paths is suppressed, leading to a damped revival at the initial hole position. In future experiments, an adiabatic evolution starting from a staggered spin pattern -- as demonstrated for the 2D square lattice~\cite{Chen2023} -- will allow us to investigate the low-energy properties of holes in the $t$-$J$~model.

In summary, we have realized 1D and 2D doped quantum magnets in a Rydberg tweezer array, implementing a fully tunable $t$-$J$-$V$ model by mapping three Rydberg states to pseudo-spins and holes~\cite{Lukas_2024_tJ}. We demonstrated the tunability of the interaction strength, 
the formation of repulsively bound holes, and how the NNN tunneling affects the perturbative tunneling of bound pairs in 1D.
Finally, we showed that the experiment can explore the 2D case, enabling us to explore the interplay between doping, spin ordering~\cite{Boninsegni2001,Harris2024,Zhang2024}, and frustration in a variety of geometries~\cite{Browaeys2020_NP}, such as ladders~\cite{Morera2024}. This demonstration of the use of three Rydberg states is a first step towards the exploration of the physics of spin-1 chains and Haldane phases~\cite{Mogerle2024}.

\newpage
%

\newpage
\subsection*{Methods}


\begin{extfig*}[t!]
    \mbox{}
    \includegraphics[width=\linewidth]{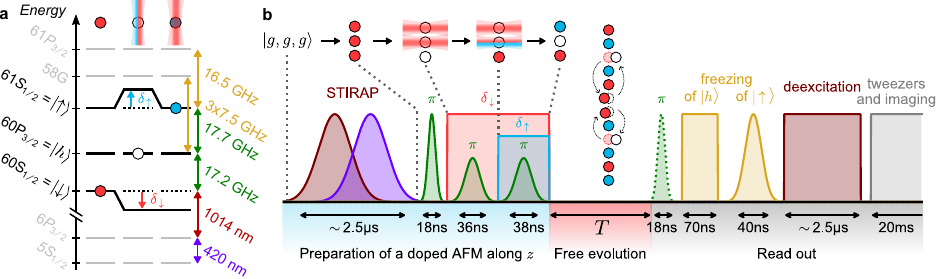}
    \caption{\label{fig:SM_sequence} {\bf Protocol for the preparation of a doped AFM along $z$.}
    \textbf{a},~Involved energy levels and transitions. The states used in the mapping to the $t$-$J$-$V$ Hamiltonian are indicated in black. Each of the three columns represents one class of atoms: non-addressed atoms (left column) are prepared in~$\ket{\downarrow}$, atoms with both $\delta_\downarrow$ and $\delta_\uparrow$ light shifts (center column) are prepared in~$\ket{h}$, and atoms with only the $\delta_\uparrow$ light shift (right column) are prepared in~$\ket{\uparrow}$.
    \textbf{b},~Experimental sequence. After Rydberg excitation using STIRAP, we apply a sequence of microwave pulses combined with the site-resolved light shifts $\delta_\uparrow$ and $\delta_\downarrow$. Then, the light shifts are switched off for a duration~$T$, during which the system evolves under the $t$-$J$-$V$ interactions. To read out the atomic states, we first perform two "freezing" pulses to stop the dynamics, that respectively act on states~$\ket{h}$ and $\ket{\uparrow}$. Then, we deexcite the atoms from~$\ket{\downarrow}$ to the ground state manifold, switch on the tweezers and image the recaptured atoms using fluorescence. An additional microwave $\pi$-pulse (dotted green line) can be used to read out the state~$\ket{h}$.
    }
    \label{fig:experimental_sequence}
\end{extfig*}

\noindent{\bf Experimental setup.} Our implementation of the $t$-$J$-$V$ model is based on the following mapping: $\ket{\downarrow}=\ket{60S_{1/2},m_J=1/2}$, $\ket{\uparrow}=\ket{61S_{1/2},m_J=1/2}$ and $\ket{h}=\ket{60P_{3/2},m_J=-1/2}$. We use a 46~G quantization magnetic field which is oriented in the atomic plane of the optical tweezers. The resonant microwave frequencies of the transitions $\ket{\downarrow}\leftrightarrow\ket{h}$ and $\ket{h}\leftrightarrow\ket{\uparrow}$ are $17179$\,MHz and $17675$\,MHz, respectively. All microwave pulses are emitted by antennas placed outside the vacuum chamber.

We use two 1014~nm lasers to generate site-resolved light shifts, respectively on the states~$\ket{\downarrow}$ and $\ket{\uparrow}$. One laser is blue-detuned by $300$~MHz from the transition between the intermediate state $\ket{6P_{3/2}}$ and $\ket{\downarrow}$, resulting in a light shift $\delta_\downarrow \sim -2\pi \times 40$~MHz on the state~$\ket{\downarrow}$; the local control is achieved by diffraction on a spatial light modulator. Another laser is red-detuned by $200$~MHz from the transition $\ket{6P_{3/2}}$ and $\ket{\uparrow}$, creating a light shift $\delta_\uparrow \sim 2\pi \times 30$~MHz on the state~$\ket{\uparrow}$; it is diffracted on an acousto-optical modulator.
\\
\\

\begin{extfig*}[t!]
    \mbox{}
    \includegraphics[width=\linewidth]{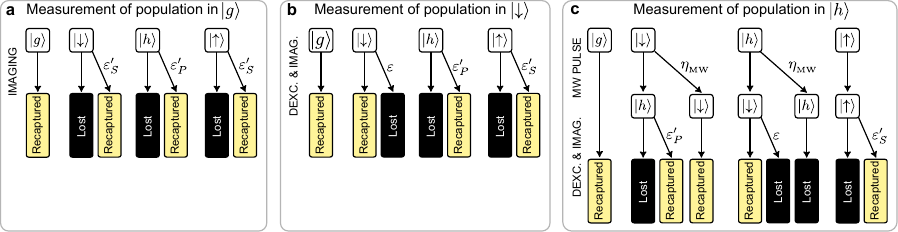}
    \caption{\label{fig:SM_error_tree} {\bf Simplified error tree associated to our detection schemes.}
    The three panels \textbf{a},\textbf{b} and \textbf{c} show the detection errors for the populations in respectively~$\ket{g}$, $\ket{\downarrow}$ and $\ket{h}$. For each scheme, we represent the probability of detection events for the four states $\ket{g}$, $\ket{\downarrow}$, $\ket{h}$ and $\ket{\uparrow}$, at first order in the detection errors $\varepsilon$, $\varepsilon_{S,P}'$ and $\eta_{_\mathrm{MW}}$.
    }
    \label{fig:error_tree}
\end{extfig*}

\noindent{\bf State initialization.}
Atoms are randomly loaded into an array of optical tweezers with a probability of $\sim 50\%$, and then rearranged by a single moving tweezer generated by a 2D AOD. After the rearrangement, we sequentially use optical molasses and Raman sideband cooling to lower the temperature of the atoms. Then the atoms are optically pumped to $\ket{g}=\ket{5S_{1/2}, F=2, m_{F}=2}$ via a $\sigma^{+}$-polarized 795~nm laser. Before the Rydberg excitation, the tweezer depth is adiabatically ramped down by a factor $\sim$~4 to reduce the momentum dispersion of the atomic wavefunctions. We then switch off the tweezers, and use a two-photon stimulated Raman adiabatic passage (STIRAP) with 420~nm and 1014~nm lasers to excite all atoms to $\ket{\downarrow}$.

Extended Data Fig.\,\ref{fig:experimental_sequence} shows a typical Rydberg sequence for the preparation of a doped AFM state along~$z$. We define three classes of sites: the sites where we want to initialize an atom in $\ket{\uparrow}$ are addressed by the $\delta_\downarrow$-laser; the ones where we want an atom in $\ket{h}$ are addressed by both the $\delta_\downarrow$- and $\delta_\uparrow$-lasers; and the ones where we want an atom in $\ket{\downarrow}$ are not addressed. First, we apply a $18$~ns microwave $\pi$-pulse to bring all the atoms from $\ket{\downarrow}$ to $\ket{h}$ (all microwave pulses have a Gaussian envelope, and their duration is given as their standard deviation multiplied by $\sqrt{2\pi}$). Second, we switch on the light shifts $\delta_\downarrow$ and apply a $36$~ns microwave $\pi$-pulse on resonance with the bare $\ket{h}\leftrightarrow\ket{\downarrow}$ transition. This transfers the non-addressed atoms from $\ket{h}$ back to $\ket{\downarrow}$, while the addressed atoms remain in $\ket{h}$. Third, we switch on the light shifts $\delta_\uparrow$ and apply a $38$~ns microwave $\pi$-pulse on resonance with the bare $\ket{h}\leftrightarrow\ket{\uparrow}$ transition. Finally, we simultaneously switch off both light shifts on $\ket{\uparrow}$ and $\ket{\downarrow}$ and let the system evolve under the $t$-$J$-$V$ Hamiltonian.

To prepare a doped FM state along~$z$, we remove the third microwave pulse and the light shifts~$\delta_\uparrow$, and we shine the light shift~$\delta_\downarrow$ on the site where we want to prepare a hole. This results in the initialization of holes ($\ket{h}$) on addressed sites, and spins down ($\ket{\downarrow}$) on non-addressed sites.

Finally, to prepare a doped magnet along~$x$, we first prepare a doped magnet along~\revise{$z$} and then apply a $\pi/2$ rotation around~$y$ between the states $\ket{\downarrow}$ and $\ket{\uparrow}$. This rotation is achieved via a two-photon microwave pulse with effective Rabi frequency $2\pi \times 9$~MHz, using an intermediate state detuning of~$\sim 200$~MHz from both $60P_{1/2}$ and $60P_{3/2}$ states.
\\
\\

\noindent{\bf State detection.}
Our readout sequence is based on the mapping of one Rydberg state to the ground state manifold $5S_{1/2}$, which is then imaged by fluorescence. For example, to measure the population in the Rydberg state $\ket{\downarrow}$, we apply a $\sim 2.5$~µs pulse of $1014$~nm light that deexcites $\ket{\downarrow}$ to the ground state manifold via the short-lived intermediate state $\ket{6P_{3/2}}$. After that, we switch the tweezers back on, in order to recapture the atoms in the ground state manifold while expelling the ones in Rydberg states via the ponderomotive force. We then image the recaptured atoms in their ground state with site-resolved fluorescence. A recaptured atom is then interpreted as $\ket{\downarrow}$; a lost atom means that the atom is in one of the two remaining states $\ket{h}$ or $\ket{\uparrow}$.

To measure the hole ($\ket{h}$) population, we add a microwave $\pi$-pulse before the deexcitation pulse, exchanging the populations of $\ket{h}$ and $\ket{\downarrow}$, and we map the recaptured atoms to $\ket{h}$.

The duration of the deexcitation is on the same order of magnitude as the typical interaction times $2\pi/J_\perp \sim 1.5$~µs. To prevent unwanted dynamics during the readout, we implement a two-step freezing protocol: (i) a three-photon transition at 7.5~GHz with a duration of 70~ns shelves the $\ket{60P_{3/2}, m_J=-1/2}$ population to $\ket{58G}$, and (ii) a microwave $\pi$-pulse transfers the atoms in $\ket{\uparrow}$ to $\ket{61P_{3/2}, m_J=3/2}$, which has negligible interaction with $\ket{\downarrow}$. The atomic transitions involved in the freezing scheme are represented as yellow arrows in Extended Data Fig.~\ref{fig:experimental_sequence}a. Those freezing pulses are applied successively just before the deexcitation.

\noindent{\bf Error budget: State preparation and detection.}
To calibrate the initial state preparation errors, we perform three measurements: one for the ground-state population~$p_g^{\mathrm{meas}}$, one for the $\ket{\downarrow}$-state population~$p_\downarrow^{\mathrm{meas}}$ and one for the $\ket{h}$-state population~$p_h^{\mathrm{meas}}$. Those measurements are performed using the sequence explained in the previous section (to measure~$p_g^{\mathrm{meas}}$, we simply remove the deexcitation pulse).

We then estimate the actual populations~$p_{g,\downarrow,h,\uparrow}^{\text {act}}$  in the states $\ket{g}, \ket{\downarrow}, \ket{h}, \ket{\uparrow}$ by correcting for detection errors (this correction is only applied to the data used in the numerical simulations, the experimental data shown in the main text still corresponds to the raw data). Our error model for detection errors assumes independent errors that depend on the measurement sequence. The error tree associated to each of the three measurement sequences is shown in Extended Data Fig.\,\ref{fig:error_tree}. To first order, the effect of the detection errors is 
\begin{align}
\left[\begin{array}{c}
    p_g^{\mathrm{meas}} \\
    p_\downarrow^{\mathrm{meas}} \\
    p_h^{\mathrm{meas}}
\end{array}\right] &=
M_\mathrm{err}
\left[\begin{array}{l}
    p_g^{\text {act }} \\
    p_\downarrow^{\text {act }} \\
    p_h^{\text {act }} \\
    p_\uparrow^{\text {act }}
\end{array}\right] \notag \\
\text{with} \; \; M_\mathrm{err} &=
\left[\begin{array}{cccc}
    1 & \varepsilon_{S}' & \varepsilon_{P}' & \varepsilon_{S}' \\
    1 & 1-\varepsilon & \varepsilon_{P}' & \varepsilon_{S}' \\
    1 & \varepsilon_{P}' & 1-\varepsilon & \varepsilon_{S}'
\end{array}\right]
\label{eq:model_detection_errors}
\end{align}
Here, $\varepsilon_S' \approx 5 \pm 1$~\% (resp. $\varepsilon_P' \approx 2 \pm 1$~\%) is the probability that an atom initially in $\ket{\downarrow}$ or in $\ket{\uparrow}$ (resp. in $\ket{h}$) decays to the ground state during the measurement, due to the finite Rydberg lifetimes; $\varepsilon \approx 2 \pm 1$~\% is the probability that an atom in $\ket{\downarrow}$ is lost before the imaging, due to the finite fidelity of the deexcitation and of the imaging. 
In Eq.~(\ref{eq:model_detection_errors}), we do not account for the error of the microwave $\pi$-pulse involved in the measurement of $p_h^{\mathrm{meas}}$ (denoted by $\eta_{_\mathrm{MW}}$ in Extended Data Fig.~\ref{fig:SM_error_tree}c), since its value strongly depends on the interactions.

To estimate the actual populations, we minimize the following cost function:
\begin{equation}
\label{eq:error-cost-function}
    \left|M_\mathrm{err}\left[\begin{array}{cc}
        p^{\text{act}}_{g}\\
        p^{\text{act}}_\uparrow\\
        p^{\text{act}}_h\\
        p^{\text{act}}_\downarrow\end{array}\right]
    -\left[\begin{array}{cc}
        p^{\text{meas}}_{g}\\
        p^{\text{meas}}_\uparrow\\
        p^{\text{meas}}_h\end{array}\right]
    \right|^2
\end{equation}
under the constraints $p^{\text{act}}_{g} + p^{\text{act}}_\uparrow + p^{\text{act}}_h + p^{\text{act}}_\downarrow = 1$, and $p^{\text{act}} \geq 0$. Here, we have neglected the residual population in Rydberg states other than $\ket{\downarrow}$, $\ket{h}$ and $\ket{\uparrow}$. We find this maximum likelihood approach to be robust against projection noise. 

The estimated state preparation errors in the initial state after correcting for the detection errors are shown in Extended Data Fig.\,\ref{fig:SM_error} in the 1D~geometry, and in Extended Data Fig.\,\ref{fig:SM_error_2d} in the 2D~geometry. We obtain preparation fidelities of $80$ to $95$~\% per site by correcting the measurements at $T=0$us. Those preparation errors arise primarily from three sources: Rydberg interactions during the pulses, STIRAP imperfections ($\sim 2$~\%), and depumping to the ground state caused by the addressing light shifts.
\\

\begin{extfig*}[t!]
    \mbox{}
    \includegraphics[width=\textwidth]{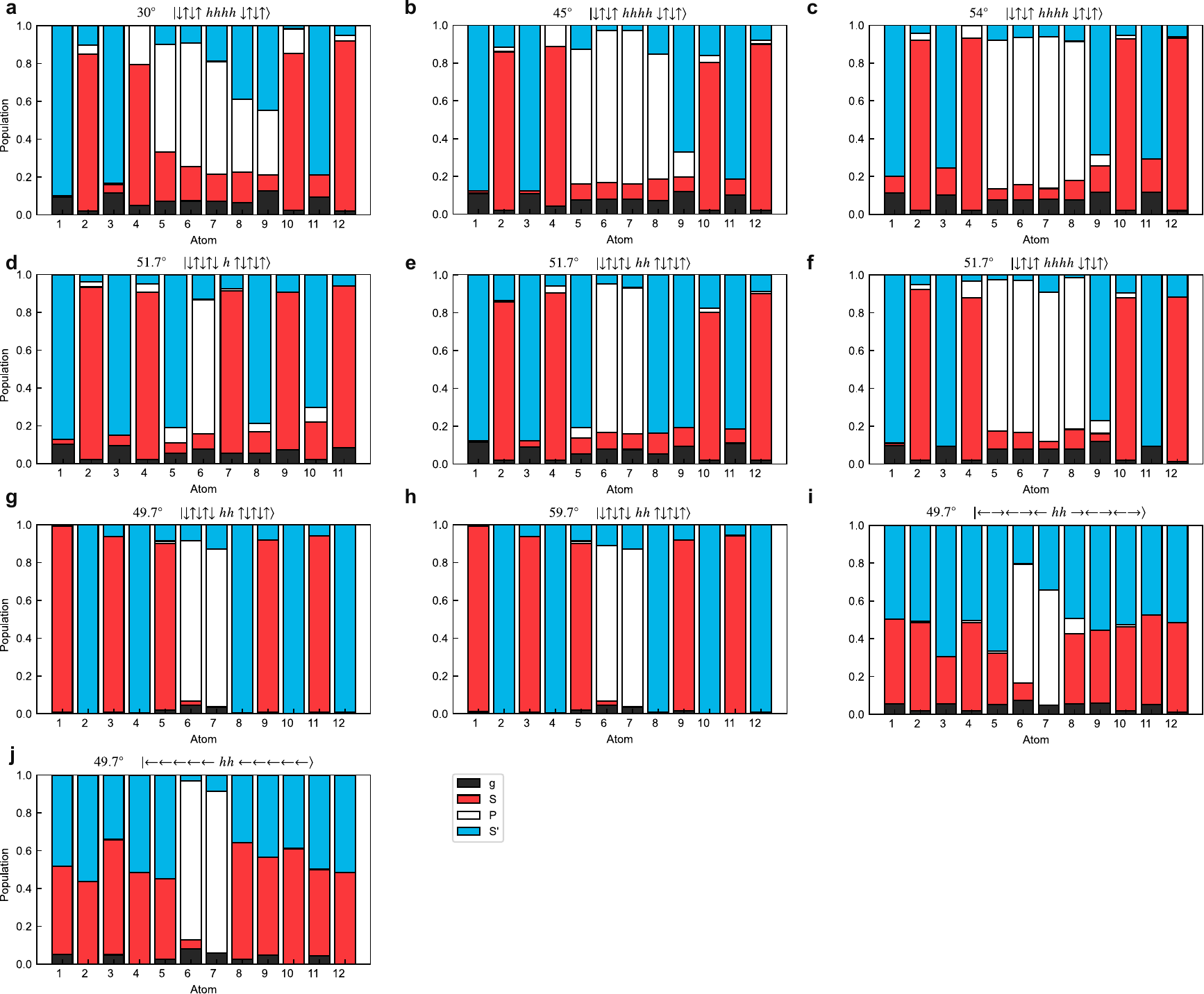}
    \caption{\label{fig:SM_error} {\bf State preparation errors for 1D chain.} Site-resolved population measurements for different initial state preparations at various angles $\theta$ relative to the quantization axis, showing the initial state preparation error. The detection error is corrected through the maximal likelihood method, using the cost function of Eq.~(\ref{eq:error-cost-function}). \textbf{a-c}, Population distributions for an initial N\'eel state with four holes in the center at $\theta=30\degree$ (a), $\theta=45\degree$ (b), and $\theta=54\degree$ (c). \textbf{d-f}, population distributions for different number of holes at $\theta=51.7\degree$, (d) 1 hole, (e) 2 holes, (f) 4 holes. 
    \textbf{g-j} Population distributions for different magnetic backgrounds: antiferromagnetic along z at $\theta=49.7\degree$ (g) and $\theta=59.7\degree$, and antiferromagnetic and ferromagnetic along x (i,j). The measured populations are shown for the four possible states: ground state (black), $\ket{\downarrow}$ (red), $\ket{h}$ (white), and $\ket{\uparrow}$ (blue). These errors are systematically included in all numerical simulations presented in the main text.}
\end{extfig*}
\begin{extfig*}[t!]
    \mbox{}
    \includegraphics[width=\textwidth]{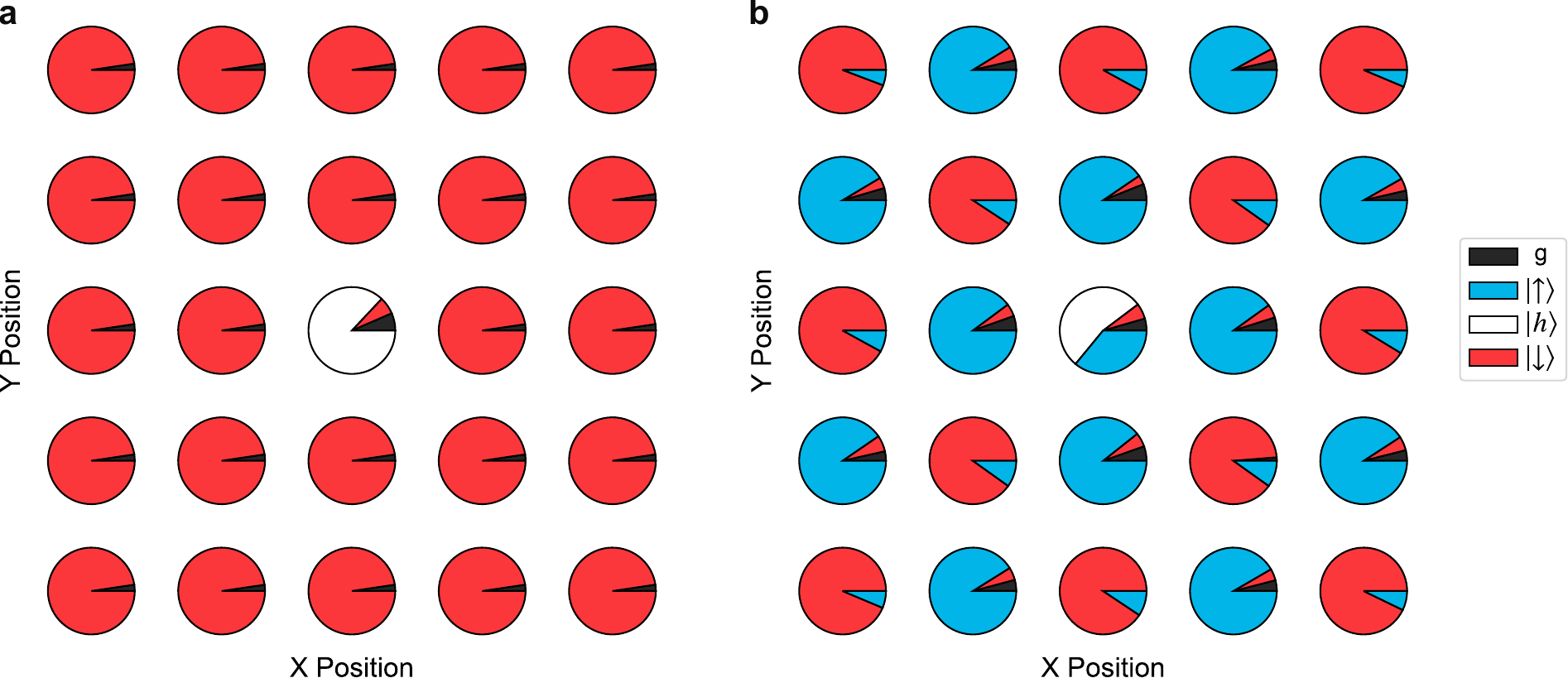}
    \caption{\label{fig:SM_error_2d} {\bf State preparation errors for 2D array.} \textbf{a}, Population measurement for a ferromagnetic state (red) with a single hole $\ket{h}$ (white) at the center. \textbf{b}, Population measurement for a N\'eel-ordered state alternating between $\ket{\uparrow}$ (blue) and $\ket{\downarrow}$ (red) with a central hole. The pie charts at each site show the relative populations of the four possible states: ground state (black), $\ket{\downarrow}$ (red), $\ket{h}$ (white), and $\ket{\uparrow}$ (blue). Detection errors have been corrected using a maximum likelihood estimation method. The observed state preparation fidelities are incorporated into the numerical simulations presented in the main text.}
\end{extfig*}

\noindent{\bf Hamiltonian mapping.} The Rydberg Hamiltonian we implement can be mapped exactly to a $t$-$J$-$V$-$W$~model, which we derive in this section.
The bare Rydberg Hamiltonian can be written as
\begin{equation} \label{eq:bare-Ryd}
\hat{H}^{\text{Ryd}} = \sum_{i<j}\hat{H}^{{\rm Ex}, C_3}_{ij} + \hat{H}^{{\rm Ex}, C_6}_{ij} + \hat{H}^{{\rm Diag}, C_6}_{ij},
\end{equation}
where~$\hat{H}^{{\rm Ex}, C_3}_{ij}$ describes the direct dipole-dipole exchange interaction, and $\hat{H}^{{\rm Ex}, C_6}_{ij}$ ($\hat{H}^{{\rm Diag}, C_6}_{ij}$) are the exchange (diagonal) components of the van-der-Waals interactions.
We express the Hamiltonian in terms of the local Rydberg basis states, i.e. an atom on site~$j$ is described by the states~$\{ \ket{60S_j}, \ket{60P_j}, \ket{61S_j} \} \equiv \{ \ket{S_j}, \ket{P_j}, \ket{S'_j} \} \equiv \{ \ket{\downarrow_j}, \ket{h_j}, \ket{\uparrow_j} \}$. The number operators at site~$j$ are thus defined as $\hat{n}^{\downarrow}_{j}=\ket{\downarrow_{j}}\bra{\downarrow_{j}}$, $\hat{n}^{h}_{j}=\ket{h_{i}}\bra{h_{j}}$, and $\hat{n}^{\uparrow}_{j}=\ket{\uparrow_{j}}\bra{\uparrow_{j}}$.

The atomic pair interactions lead to Hamiltonian~\eqref{eq:bare-Ryd}, which is given by
\begin{align}
    \hat{H}^{{\rm Ex}, C_3}_{ij} = &-\frac{t_\downarrow (\theta_{ij})}{r_{ij}^3}\big( \ket{\downarrow_i,h_j}\bra{h_i,\downarrow_j} +\mathrm{h.c.} \big) \\
    &-\frac{t_\uparrow(\theta_{ij})}{r_{ij}^3}\big( \ket{\uparrow_i,h_j}\bra{h_i,\uparrow_j} +\mathrm{h.c.} \big)
\end{align}
and 
\begin{align}
    \hat{H}^{{\rm Ex}, C_6}_{ij} &= \frac{J_\perp(\theta_{ij})}{2} \cdot \frac{1}{r_{ij}^6}\big( \ket{\downarrow_i,\uparrow_j}\bra{\uparrow_i,\downarrow_j} +\mathrm{h.c.} \big)
\end{align}
and 
\begin{align} \label{eq:Rydberg_Ham_diag}
    \hat{H}^{{\rm Diag}, C_6}_{ij} = \sum_{\alpha,\beta \in \{\downarrow,h,\uparrow\}}\frac{V_{\alpha \beta}(\theta_{ij})}{r_{ij}^6}\hat{n}^\alpha_{i}\hat{n}^\beta_{j},
\end{align}
where~$r_{ij}$ is the vector connecting atoms at site~$i$ and~$j$, and $\theta_{ij}$ the angle between the vector~$r_{ij}$ and the quantization axis defined by the magnetic field with magnitude~$B=46\,\mathrm{G}$.
The derivation shows that the direct dipole-dipole and the van-der-Waals exchange interactions are directly related to the tunneling~$t$ and spin flip-flop~$J_\perp$, see Extended Data Table~\ref{table:bosonic-tJ-Ryd-}. We emphasize that our choice of Rydberg states only have a small anisotropy in the tunneling amplitudes~$t_\uparrow (\theta_{ij}) \approx 0.95 \cdot t_\downarrow(\theta_{ij})$.

\begin{table*}[t!]
\begin{center}
\renewcommand{\arraystretch}{2.}
\newcolumntype{Y}{>{\centering\arraybackslash}X}
\begin{tabularx}{\linewidth}{|c|| Y |} 
 \hline
  $t$-$J$-$V$ & Rydberg \\
  \hline \hline
  $t_\downarrow(\theta_{ij})$ & $-C_3^{S P}(1-3\cos^2{\theta_{ij}})$ \\
  \hline
  $t_\uparrow(\theta_{ij})$ & $-C_3^{S' P}(1-3\cos^2{\theta_{ij}})$ \\
  \hline
  $J_\perp(\theta_{ij})$  & $2 \cdot C_{6,\perp}^{SS'}(\theta_{ij})$ \\
  \hline
  $J_z(\theta_{ij})$ & $C_6^{SS}(\theta_{ij}) + C_6^{S'S'}(\theta_{ij}) - 2C_6^{SS'}(\theta_{ij})$ \\
  \hline
  $V(\theta_{ij})$  & $\frac{1}{4}\left[C_6^{SS}(\theta_{ij}) + C_6^{S'S'}(\theta_{ij}) + 2C_6^{SS'}(\theta_{ij}) \right] - C_6^{SP}(\theta_{ij}) - C_6^{S'P}(\theta_{ij}) +C_6^{PP}(\theta_{ij})$ \\
  \hline
  $W(\theta_{ij})$ & $C_6^{S'P}(\theta_{ij}) - C_6^{SP}(\theta_{ij}) + \frac{1}{2}\left[C_6^{SS}(\theta_{ij}) - C_6^{S'S'}(\theta_{ij}) \right]$ \\
  \hline
  $h_j^z(\theta_{ij})$ & $\frac{1}{2}\sum_{i} \frac{1}{r^6_{ij}}\left[  C_6^{S'S'}(\theta_{ij}) - C_6^{SS}(\theta_{ij}) \right]$ \\
  \hline
  $\mu_j(\theta_{ij})$ & $\frac{1}{4}\sum_{i} \frac{1}{r^6_{ij}}\left[  C_6^{SS}(\theta_{ij}) + C_6^{S'S'}(\theta_{ij}) + 2C_6^{SS'}(\theta_{ij}) - 2C_6^{SP}(\theta_{ij}) - 2C_6^{S'P}(\theta_{ij}) \right]$\\
  \hline
\end{tabularx}
\caption{\label{table:bosonic-tJ-Ryd-} \textbf{Mapping of interaction strengths.} The coupling strength of the $t$-$J$-$V$~model, see Eq.~\eqref{_Hamiltonian} and Eq.~\eqref{eq:bare-Ryd}, are related to the atomic pair interactions.}

\end{center}
\end{table*}

\begin{table*}[htbp]
\begin{center}
\renewcommand{\arraystretch}{2.0}
\begin{tabular}{|c||c|c|c|c|c|c|c|c|c|} 
 \hline
 $\theta$ (\si{\degree}) & a (\si{\micro m}) & $t_\downarrow$ (MHz) & $t_\uparrow$ (MHz) & $J_\perp$ (MHz) & $J_z$ (MHz) & $V$ (MHz) & $W$ (MHz) & $h^z$ (MHz) & $\mu$ (MHz) \\ 
 \hline \hline 
 30   & \multirow{6}{*}{\centering 9.9} & -1.10  & -1.05  & 0.57 & -0.33 & -0.10 & -0.01 & 0.02 & 0.14 \\ \cline{1-1}\cline{3-10} 
 45   &                                 & -0.43  & -0.41  & 0.64 & -0.40 &  0.72 &  0.03 & 0.02 & 0.53 \\ \cline{1-1}\cline{3-10}
 49.7 &                                 & -0.22  & -0.21  & 0.67 & -0.42 &  0.81 &  0.03 & 0.02 & 0.58 \\ \cline{1-1}\cline{3-10} 
 51.7 &                                 & -0.13  & -0.13  & 0.68 & -0.43 &  0.82 &  0.03 & 0.02 & 0.58 \\ \cline{1-1}\cline{3-10} 
 54.7 &                                 &  0.03 &  0.02 & 0.70 & -0.45 &  0.82 &  0.03 & 0.02 & 0.58 \\ \cline{1-1}\cline{3-10} 
 59.7 &                                 &  0.21  &  0.20  & 0.72 & -0.47 &  0.74 &  0.02 & 0.02 & 0.54 \\ \hline 
 90   & 12                              &  0.51  &  0.48  & 0.26 & -0.18 & -0.02 & -0.00 & 0.01 & 0.05 \\ \hline 
\end{tabular}
\caption{\label{table:tJVW_interactions} \textbf{Interaction strengths of the bosonic $t$-$J$-$V$ model}. Calculated values of the interaction strength for the experimental configurations used in the main text, and the full angle dependence can be found in Fig.~\ref{fig:couplings}b. The calculated values indicate that $W, h_z$ are typically negligible compared to other energy scales. While the chemical potential $\mu$ can be significant, its uniformity in the bulk of a regular lattice means its primary physical influence is expected at the system boundaries.
}
\end{center}
\end{table*}

Next, we use the hard-core constraint of the Rydberg occupation number, i.e.
\begin{align}
    \hat{n}^{\downarrow}_j + \hat{n}^{\uparrow}_j + \hat{n}^{h}_j = 1 ~~~\forall j,
\end{align}
to express the spin-1/2 operators as
\begin{align}
\begin{split} \label{eq:spin-number-identity}
    \hat{n}^{\downarrow}_j = -\hat{S}^z_j + \frac{1}{2} - \frac{1}{2} \hat{n}^{h}_j \\
    \hat{n}^{\uparrow}_j = +\hat{S}^z_j + \frac{1}{2} - \frac{1}{2} \hat{n}^{h}_j.
\end{split}
\end{align}
Inserting Eqs.~\eqref{eq:spin-number-identity} into the diagonal van-der-Waals Hamiltonian~\eqref{eq:Rydberg_Ham_diag} yields
\begin{align}
\begin{split} \label{eq:spin-Hamiltonian_diag}
    \hat{H}^{{\rm Diag},C_6}_{ij} &= \frac{J^z(\theta_{ij})}{r^6_{ij}} \hat{S}^z_i\hat{S}^z_j + \frac{V(\theta_{ij})}{r^6_{ij}}\hat{n}^{h}_i \hat{n}^{h}_j \\
    &+ \frac{W(\theta_{ij})}{r^6_{ij}}\left( \hat{S}^z_i \hat{n}^{h}_j + \hat{n}^{h}_i\hat{S}^z_j \right)\\
    &+ \frac{h^z_j(\theta_{ij})}{r^6_{ij}}\hat{S}^z_j - \frac{\mu_j(\theta_{ij})}{r^6_{ij}}\hat{n}^{h}_j \\
    &+ \mathrm{const.}
\end{split}
\end{align}
A few of the above terms are neglected in the main text. First, for the chosen atomic levels, the spin-hole interaction $W$ is negligible compared with the other interaction strengths such as the hole-hole interaction~$V$. Second, the field terms~$\propto h^z_j$ and~$\mu_j$ are mostly flat in the bulk but are different on the boundary, where an atom in a 1D chain has only one instead of two nearest neighbors. The relations between the couplings in the $t$-$J$-$V$~model and the Rydberg Hamiltonian~\eqref{eq:bare-Ryd} are summarized in Table~\ref{table:bosonic-tJ-Ryd-}.

To obtain the coupling amplitudes of the $t$-$J$-$V$~model, see Eq.~\eqref{_Hamiltonian}, we calculate the atomic pair interactions at magnetic field magnitude~$B=46\,\mathrm{G}$ for angles $\theta_{ij} \in [0,\pi]$ using the \textit{pairinteraction} software package~\cite{Weber2017}, see solid markers in Extended Data Fig.\,\ref{fig:couplings}a.
The angular dependence of the van-der-Waals interactions~$\propto C^{\alpha\beta}_6 (\theta_{ij})$ can be well fitted by the function $f^{\alpha\beta}(\theta_{ij}) = F^{\alpha\beta}_1 + F^{\alpha\beta}_2 \cos^2\theta_{ij} + F^{\alpha\beta}_3 \cos^4\theta_{ij}$, where $F^{\alpha\beta}_n$ are fit parameters that depend on the quantum numbers~$\alpha,\beta \in \{S,S',P\}$ and magnitude of the magnetic field. To be explicit, we fit each interaction coefficient $C^{\alpha\beta}_6 (\theta_{ij})$ to the function $f^{\alpha\beta}$, see solid lines in Extended Data Fig.\,\ref{fig:couplings}a. We use the extracted $C^{\alpha\beta}_6 (\theta_{ij})$ and the calculated $C^{\alpha\beta}_3$ coefficients to determine the interaction strengths at an atomic distance of $a=9.9\,\mu\mathrm{m}$ (1D) and $a=12\,\mu\mathrm{m}$ (2D) in our numerical simulations; in Fig.\,\ref{fig:setup}d-e, we show a comparison of the theoretically calculated and experimentally measured interaction strengths.


\begin{extfig}[t!]
    \mbox{}
    \includegraphics[width=0.99\linewidth]{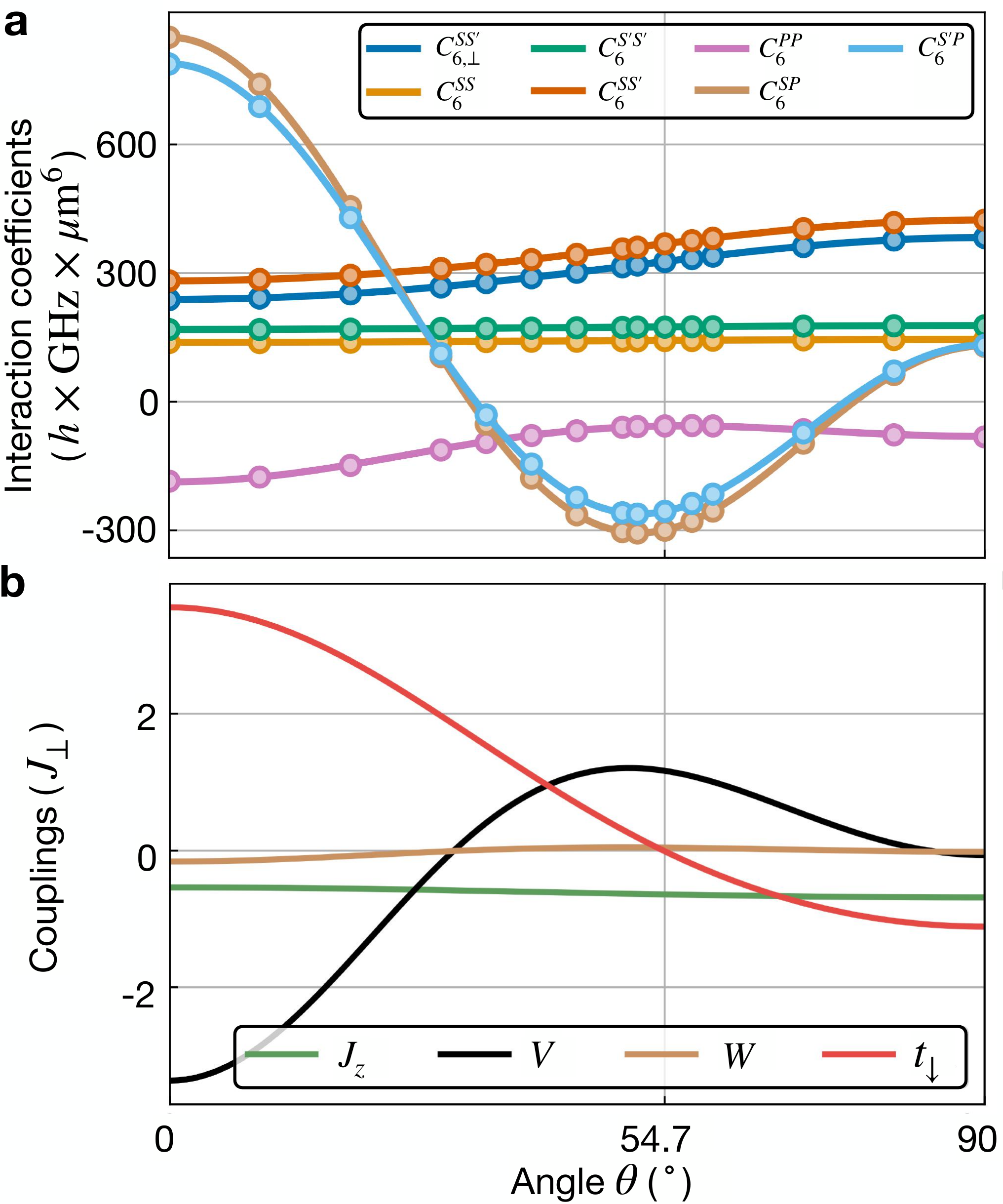}
    \caption{\label{fig:couplings} {\bf Angular dependence of interactions.} \textbf{a} The van-der-Waals interaction coefficients~$C_6^{\alpha\beta}(\theta_{ij})$ are calculated from the Rydberg pair interactions at a magnetic field magnitude of~$B=50\,\mathrm{G}$ for various angles~$\theta_{ij}$ between the inter-atomic axis and the magnetic field (solid markers). We fit the each coefficient to the function~$f^{\alpha \beta}(\theta_{ij}) = F^{\alpha\beta}_1 + F^{\alpha\beta}_2 \cos^2\theta_{ij} + F^{\alpha\beta}_3 \cos^4\theta_{ij}$ (solid line). \textbf{b} The coupling amplitudes of the $t$-$J$-$V$-$W$ model are computed using Table~\ref{table:bosonic-tJ-Ryd-}. The spin-hole interaction~$W/J_\perp \approx 0$ for all angle and hence we can neglect it for all practical purposes in our interpretation of the experimental results. Note that numerical calculations are based on the full Rydberg Hamiltonian.
    }
\end{extfig}

\begin{extfig}[t!]
    \mbox{}
    \centering
    \includegraphics[width=0.9\linewidth]{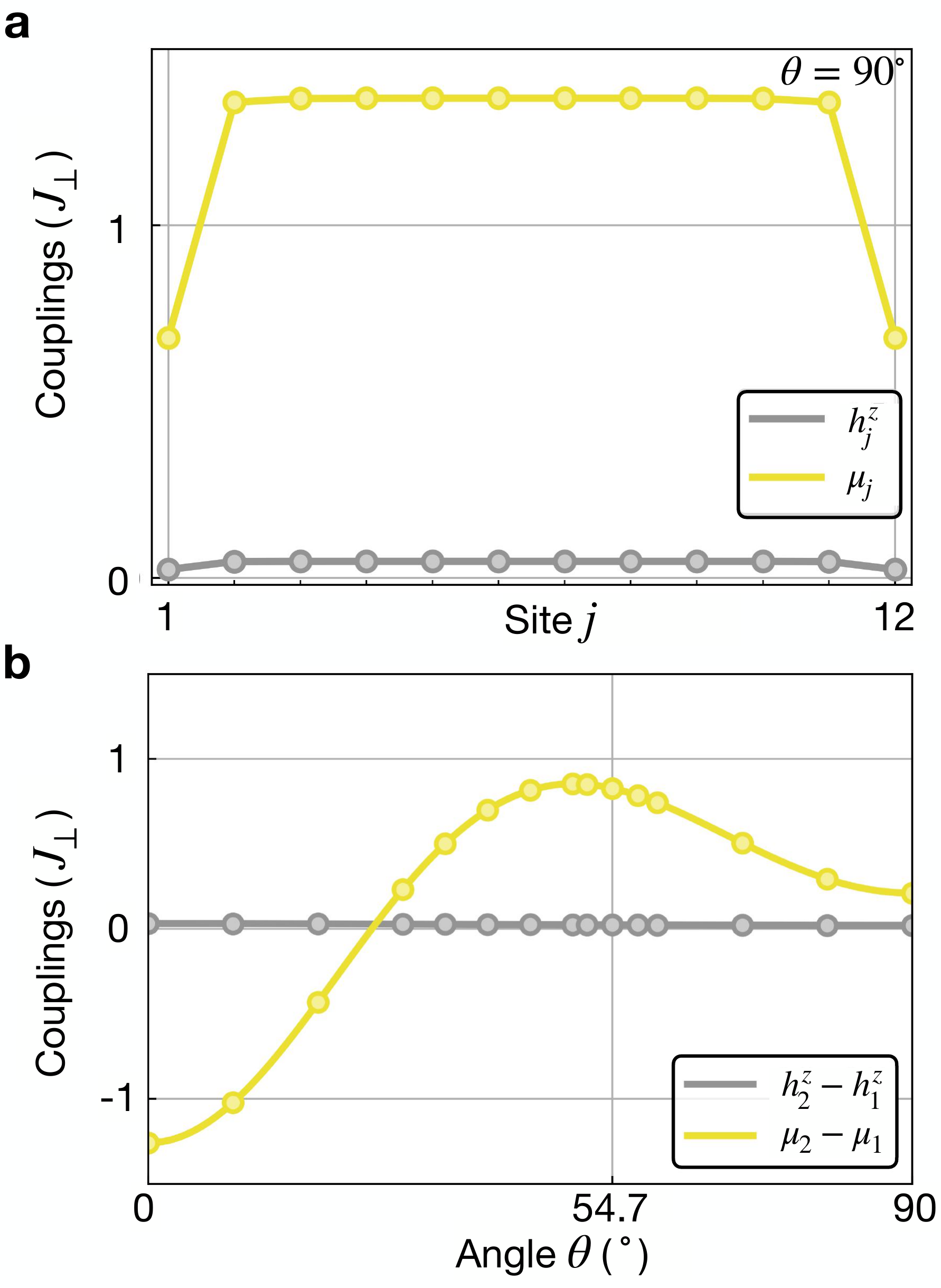}
    \caption{\label{fig:boundary} {\bf Boundary terms.} Site-dependent field terms~$h^z_j$ and~$\mu_j$ calculated as in Fig.~\ref{fig:couplings}. \textbf{a} Fields at angle~$\theta=90^\circ$ for the 12-site chain realized in the experiment. As the difference in magnitude between the fields at the boundary compared to the bulk is large, those terms can play a role in the presence of boundaries. \textbf{b} To quantify the boundary terms, we plot the difference of the fields between site $2$ and $1$ as a function of the angle~$\theta$.} 
\end{extfig}

The angular dependence of the $C^{\alpha\beta}_6 (\theta_{ij})$ and $C_3(\theta_{ij})$ coefficients lead to a strong angular dependence of the couplings in the $t$-$J$-$V$~model, see Extended Data Table~\ref{table:bosonic-tJ-Ryd-}. This allows us to tune through a wide parameter regime in the 1D~chain by changing the angle between the quantization axis and the unit vector along the chain. In Extended Data Fig.\,\ref{fig:couplings}b, the interactions $t_\downarrow(\theta_{ij})$, $J_z(\theta_{ij})$, $V(\theta_{ij})$ and~$W(\theta_{ij})$ at lattice spacing $a=9.9\,\mu\mathrm{m}$ are plotted in units of the spin flip-flop interaction~$J_\perp(\theta_{ij})$; hence the absolute interaction scale changes from $J_\perp(0) = 2\pi \times 507\,\mathrm{kHz}$ to $J_\perp(90\degree) = 2\pi \times 813\,\mathrm{kHz}$.

We note that the spin-hole interaction~$\propto W(\theta_{ij})$ is negligible for the chosen Rydberg states. Thus, we do not expect it to play a role on the timescales of the performed experiments. Nevertheless, since we use the bare Rydberg Hamiltonian for our numerical simulations, all terms including the boundary terms in Hamiltonian~\eqref{eq:spin-Hamiltonian_diag} are fully taken into account. We have confirmed the mapping between the bare Rydberg model to the $t$-$J$-$V$~model according to Extended Data Table~\ref{table:bosonic-tJ-Ryd-} by comparing the full spectra in a small system.

Last, we consider the boundary and field terms that appear in the mapping from Rydberg states to (doped) spin models. They result, in pure spin models, in magnetic field terms~$h^z_j$; in our model we obtain an additional chemical potential term~$\mu_j$. In Extended Data Fig.~\ref{fig:boundary}a, we show the spatial dependence of the field terms for the~$L=12$ site chain. In the bulk, the field terms are approximately constant and, due to the particle and magnetization conservation  in our model, the field term results in a constant energy shift. However, in the presence of boundaries, where an atom has only one instead of two NN, the field terms are approximately half in magnitude compared to the bulk; hence these terms only have to be considered at the boundary. In Extended Data Fig.~\ref{fig:boundary}b, we further study the difference of the couplings between site $2$ and $1$ as we vary the angle~$\theta$.
In the experiments conducted here, we do not expect the boundary terms to play a significant role in the physics. For instance, the pair dynamics predominately happens in the bulk. In future experiments with adiabatic ground state preparation, the boundaries would have to be considered.
\\
\\
\noindent{\bf Dynamical phase separation.}
Ground-state phase separation into hole-rich and hole-free regions have been studied extensively in the fermionic $t$-$J$~models, particularly in the early years of high-Tc superconductivity~\cite{emery1990separation,Marder1990,Bobroff2002}, due to its notion of self-binding of charge carriers. Also, sign-problem free quantum Monte Carlo studies report phase separation in (partially) antiferromagnetic bosonic $t$-$J$~models~\cite{Boninsegni2001}. The phase separation is found for small values of $|t/J| \ll 1$, where the kinetic energy is small enough such that the particles cannot escape from the self-bound state. Consequently, in quantum simulation platforms with superexchange-based magnetic interactions, e.g. optical lattices, the accessible parameter range does not allow one to probe phase separation.



While phase separation has previously been studied as a property of the ground state, here we develop a method to dynamically probe phase separation by a quench starting from a product state. We define states with a hole-rich connected-region~$R$ as $\ket{\psi}_{R} = \ket{\psi_s} \otimes \prod_{j \in R}\ket{h}_j$, i.e. all $N_h$~holes in the system form a cluster of volume $V(R)=N_h$ and the state~$\ket{\psi_s}$ describes the spins located at sites~$j \notin R$. To characterize the states~$\ket{\psi}_{R}$, we consider the projector
\begin{align}
\hat{\mathcal{P}}_{N_h}^{h}=\sum_R\prod_{j \in R} \hat{n}_j^h
\end{align}
such that states the expectation values $\mathcal{P}_{N_h}^{h} = \langle \hat{\mathcal{P}}_{N_h}^{h} \rangle$ describes the overlap with hole-rich connected-regions. For example both states $\ket{\uparrow\downarrow\uparrow\downarrow hhhh\uparrow\downarrow\uparrow\downarrow}$ and $\ket{\downarrow\uparrow\uparrow\uparrow\downarrow \downarrow hhhh\uparrow\downarrow}$ are assigned the same values~$\mathcal{P}_{4}^{h}=1$ and are therefore labeled as perfectly phase separated by our definition.

Our ansatz is independent of the spin state in the hole-free regions, which is crucial in our analysis since we study far-from-equilibrium states with an energy density well-below the highest-energy state; note that the hole-hole repulsion stabilizes phase separation for negative temperature states.
In practice, the initial state~$\ket{\psi_0}=\ket{\uparrow\downarrow\uparrow\downarrow hhhh\uparrow\downarrow\uparrow\downarrow}$ experiences strong dynamics in the spin sector~$\ket{\psi_s}$, while the hole-rich sector remains stable throughout the experiment with some fluctuations at its boundary. Hence, the hole and spin sector appear separated under dynamics.

To quantify whether an initial state $\ket{\psi_0}$ remains phase separated under time evolution, we perform exact diagonalization on the $L=12$ chain to obtain the entire dynamically accessible spectrum, i.e. we fix total magnetization~$S_{\rm{tot}}^{z}=0$ and the number of holes $N_{h}=4$. 
In particular, we calculate the expectation value $\bra{n} \hat{\mathcal{P}}_{4}^{h} \ket{n}$ for all eigenstates~$\ket{n}$, and we define the weighted diagonal ensemble
\begin{align}
\langle \hat{\mathcal{P}}_{4}^{h} \rangle_{\mathrm{w}} = \sum_n |\langle n | \psi_0 \rangle |^2 \bra{n} \hat{\mathcal{P}}_{4}^{h} \ket{n},
\end{align}
which we use as an order parameter for dynamical phase separation for the triplet ($\hat{H}$,\,$\hat{\mathcal{P}}_{4}^{h}$,\,$\ket{\psi_0}$).
In fact, the weighted diagonal ensemble is directly related to the infinite time expectation value $\mathcal{P}_{4}^{h}(t \rightarrow \infty)$ assuming eigenstate thermalization hypothesis:
\begin{align}
\begin{split}
    \mathcal{P}_{4}^{h}(t) &:= \bra{\psi(t)} \hat{\mathcal{P}}_{4}^{h} \ket{\psi(t)} \\ &= \sum_{n,m} \langle \psi_0 | n \rangle \bra{n} \hat{\mathcal{P}}_{4}^{h} \ket{m}\langle m | \psi_0 \rangle e^{-i(E_m - E_n)t}  \\
    &~\overset{t\rightarrow \infty}{\rightarrow}  \langle \hat{\mathcal{P}}_{4}^{h} \rangle_{\mathrm{w}}.
\end{split}
\end{align}
The weighted overlap $\langle \hat{\mathcal{P}}_{4}^{h} \rangle_{\mathrm{w}}$ is plotted in Fig.\,\ref{fig:dynamics}d for a hypothetical scan of $t/J_\perp$ and $V/J_\perp$ showing a region in parameter space (dark blue) for which the initial state~$\ket{\psi_0}$ equilibrates to a state with a strong character of phase separation. As expected, the hole-hole repulsion~$V$ gives rise to a fan-like region around~$|t| = 0$ increasing with~$V$.
\\
\\
\noindent{\bf Influence of magnetic background on hole pair.}
In the limit of small tunneling~$|t| \ll |J_\perp|,|J_z|,|V|$ studied in our experiments, the qualitative behavior of the repulsively bound hole pairs can be understood by a perturbative analysis. In fact, the spin background influences both the binding energy $E_b$ of the pair together with the effective mass~$m_{\mathrm{eff}} = 1/2t_{\mathrm{eff}}$. In addition, the interference of the effective pair tunneling with the next-nearest neighbor tunneling process~$(\propto \pm t/8)$ allows us to tune the pair's mass by varying the angle~$\theta$.

The quench protocol we apply begins from an initial product state at time~$T=0$. At short times, we assume the kinetic motion of holes to be frozen, while the spin background evolves under the Hamiltonian~\eqref{_Hamiltonian} allowing us to treat the spin sector independently from the holes in the limit $|J_\perp|, |J_z| \gg |t|$. Under unitary time evolution of a translationally invariant initial spin background, the energy per bond~$\epsilon$ is conserved and for the four different initial product states given by
\begin{align}
\begin{split}
    &\epsilon_{\mathrm{zAFM}}=|J_z|/4 ~~~~~~~~~~~~ \epsilon_{\mathrm{zFM}}=-|J_z|/4 \\
    &\epsilon_{\mathrm{xAFM}}=-J_\perp/4 ~~~~~~~~~~\epsilon_{\mathrm{xFM}}=J_\perp/4,
\end{split}
\end{align}
where we neglect boundary terms.

\begin{extfig}[t!]
    \mbox{}
    \includegraphics[width=\linewidth]{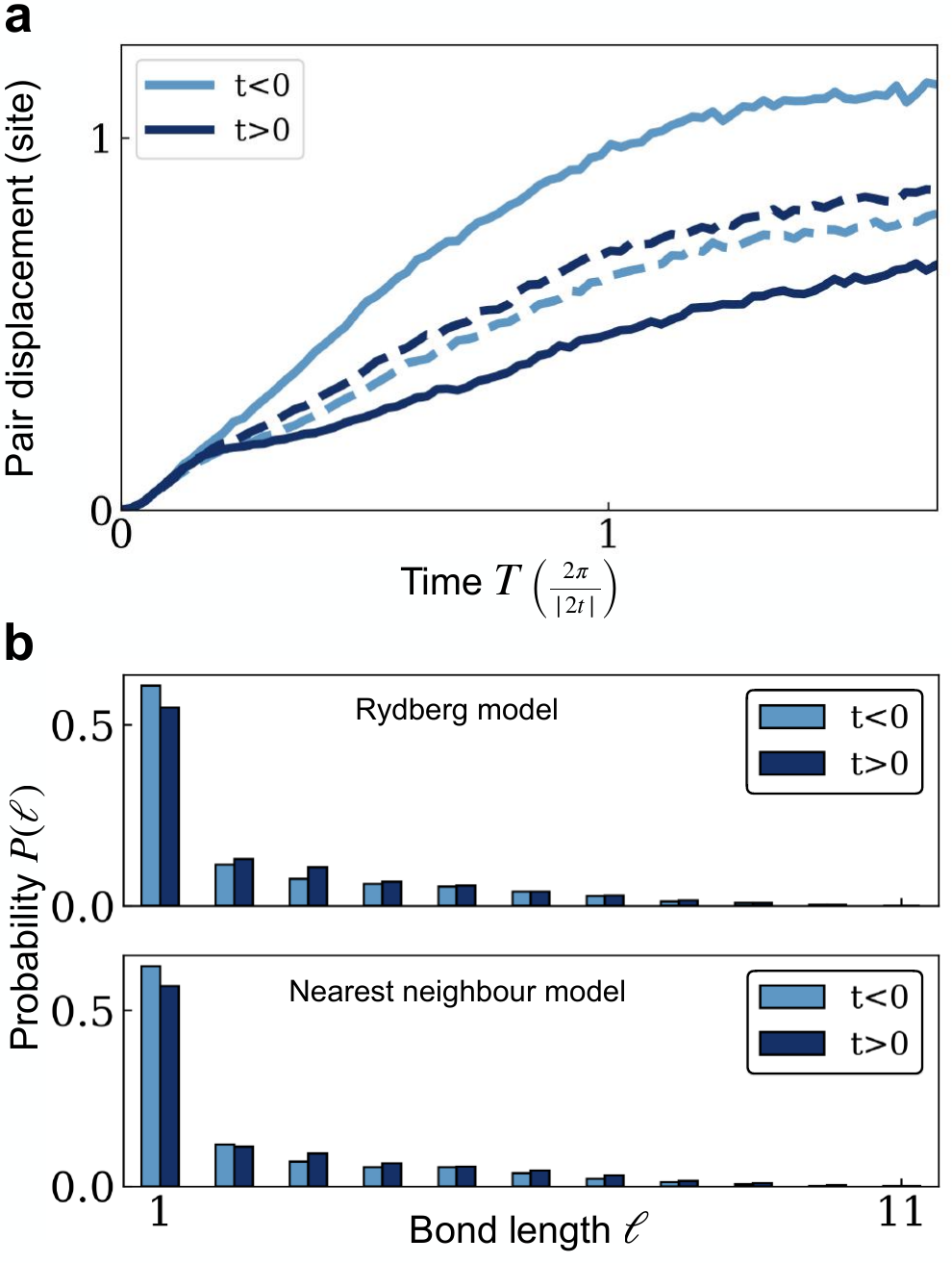}
    \caption{\label{fig:pair_dist_SR_ZAFM} {\bf Interference from NNN tunnelings~$t'$.} We numerically study the time evolution of an initial hole pair in the $L=12$ \mbox{z-AFM} without errors under (i) the full Rydberg Hamiltonian and (ii) a model with interactions truncated beyond the NN couplings. The angles are chosen to be~$\theta=49.7\degree$ ($t<0$) and ~$\theta=59.7\degree$ ($t>0$), see Fig.\,\ref{fig:perturbative_tunneling}. \textbf{a} We plot the center-of-mass pair displacement of the pair. The solid line corresponds to model (i) and corresponds to an ideal experiment without errors; the difference in the center-of-mass pair displacement is clearly visible, which we explain by the interference between perturbative and NNN tunneling~$t'$. When we consider the model with only NN couplings (ii), shown as dashed lines, the strong difference in the pair displacement disappears. \textbf{b} We compare the pair distance histograms at time~$T=1.6\times\frac{2\pi}{|2t|}$ and find that the hole is more tightly bound for $\theta=49.7\degree$ than for $\theta=59.7\degree$ independent of the range of tunnelings. This is consistent with our perturbative description of the pair, where the binding energy only depends on the hole-hole repulsion~$V$ and spin interaction~$J_z$.
    }
\end{extfig}

Next, we consider the perturbative tunneling of a hole pair immersed into the thermalized spin background; here we define the pair to be constituted by neighboring holes. The energy of the initial state is composed by the energy per bond~$\epsilon$ and the hole-hole repulsion~$V$. In the tightly-bound regime, the effective perturbative tunneling of the pair can be described by a second-order process through a virtual, off-resonant state as shown in Fig.\,\ref{fig:perturbative_tunneling}a. The energy difference is then determined by $\Delta E = E_{\mathrm{init}} - E_{\mathrm{virtual}}$ and depends on the spin background:
\begin{align}
\begin{split}
    &\Delta E_{\mathrm{zAFM}}=V + \frac{|J_z|}{4} ~~~~~~~~~~~~ \Delta E_{\mathrm{zFM}}=V - \frac{|J_z|}{4} \\
    &\Delta E_{\mathrm{xAFM}}=V - \frac{J_\perp}{4} ~~~~~~~~~~\Delta E_{\mathrm{xFM}}=V + \frac{J_\perp}{4}.
\end{split}
\end{align}
The coupling between the initial and virtual states as well as from the virtual to the final state scales with the bare tunneling~$t$, and further the perturbative description includes a factor~$\chi$ associated with the finite overlap between states in the spin sector. In general, the factor~$\chi = \chi(T)$ is spin-background-dependent and time-dependent with the timescale corresponding to the equilibration time of the spin background. At late times, we find the effective perturbative tunneling~$-t^{(\mathrm{pert})} =  \chi t^2/\Delta E > 0$.

In addition to the perturbative pair tunneling process, the long-range dipolar tunnelings induce a direct tunneling of amplitude~$-t^{(\mathrm{dir})} = -t/8$, where one hole traverses the other. In our 1D scheme, the sign of the tunneling~$t$ depends on the angle~$\theta$ relative to the magic angle~$\theta_m = 54.7\degree$. The resulting total effective pair mass is then given by~$t^{(\mathrm{eff})} = t^{(\mathrm{pert})} + t^{(\mathrm{dir})}$, which we probe by the center-of-mass spreading of the hole pairs, see Fig.\,\ref{fig:perturbative_tunneling}e and Fig.\,\ref{fig:spin_background}b. In particular, in the analysis shown Fig.\,\ref{fig:spin_background}, we postselect on snapshots in the~$\ket{P}$-basis with $N_h=2$ holes and bond length~$\ell \leq 2$, from which we extract the center-of-mass spreading of the pair; the qualitative analysis is not dependent on our choice of the pair's size.


As a consistency check of our perturbative pair model, we perform a numerical analysis of the protocol described in Fig.\,\ref{fig:perturbative_tunneling}, and compare it to a model with couplings between only NN sites. First, we predict the NN model to not experience interference with a direct tunneling term, leaving only the bare perturbative tunneling of the pair, see Extended Data Fig.\,\ref{fig:pair_dist_SR_ZAFM}a. In contrast, the binding energy is not affected by the absence of the dipolar tails and we expect a difference in the binding strength in both cases, see Extended Data Fig.\,\ref{fig:pair_dist_SR_ZAFM}b. Our predictions derived from perturbation theory are consistent with the numerical simulations corroborating the influence of dipolar tunnelings present in our experiment.

\noindent{\bf Numerical simulations.}
The numerical time-evolution simulations presented in the main text and Methods, except Fig.\,\ref{fig:2d}c and d, are based on exact diagonalization methods of the Rydberg Hamiltonian, Eq.~\eqref{eq:bare-Ryd}, including interactions up to distance~$r_{ij} \leq 3$ (in units of lattice sites). We develop an error model to describe the experimental situation, which we introduce in the following in decreasing order of relevance.

{\it State preparation errors - }
Initial state preparation errors are most detrimental to our quench protocol. In particular, atoms with an initial local target state can falsely be prepared in any of the other two Rydberg states, or remain in the (non-interacting) ground state.

To model these errors, we sample from a probability distribution, initialize the (falsely prepared) initial product state, and time evolve under the Rydberg Hamiltonian. Depending on the errors and the observables, we sample between $500$ and $4000$ initial states and take one snapshot in the $\ket{S}$, $\ket{P}$ and $\ket{S'}$ basis after time~$T$. Atoms that remained in the ground state are recaptured and show up as a signal in the snapshot. 

The measurements of the bound state population~$P(\ell)$ shown in Fig.\,\ref{fig:perturbative_tunneling}d and\,\ref{fig:spin_background}a includes differences between correlation functions, which are not captured accurately by our error model. To improve the theoretical model, we have checked that including correlated errors in the initial state allows us to obtain better agreement between the numerical simulation and the experimental data. For consistency, we only present simulations underlying the same initial state error model. In the future, it is crucial to develop an error scheme that captures correlated errors to quantitatively compare higher-order correlation functions. 


{\it Detection errors - } In the experiment an atom can be falsely detected, e.g. during the deexcitation pulse of the $\ket{\downarrow}$ state, an atom in state~$\ket{h}$ can decay. This atom is then recaptured and imaged. In the numerical simulations, we include the detection errors by postprocessing snapshots, i.e. we randomly flip the bits in a snapshot according to the scheme shown in Extended Data Fig.~\ref{fig:error_tree} and the probabilities in Eq.~\eqref{eq:model_detection_errors}; this procedure requires knowledge about the actual state.

{\it Rydberg lifetime - }
We use quantum trajectory methods to include the finite lifetime of Rydberg states~\cite{Daley2014}. For each atomic state we assume two decay channels: 
\begin{enumerate}
    \item A Rydberg state decays down to the atomic ground state and appears as a signal in the snapshots. We approximate this decay rates by the value of the radiative rates (at temperature 0\,K):
    \begin{align}
    \begin{split}
        \Gamma_{60S}\approx\Gamma_{61S} &\approx (260\,\mu\mathrm{s})^{-1} \\
        \Gamma_{60P}&\approx(472\,\mu\mathrm{s})^{-1}.
    \end{split}
    \label{eq:gamma_0K}
    \end{align} 

    \item A Rydberg state has a blackbody-induced transition to other Rydberg states and is lost from the simulation. The decay rates are given by:
    \begin{align}
    \begin{split}
        \Gamma_{60S}^{\rm BB}\approx\Gamma_{61S}^{\rm BB} &\approx (157\,\mu\mathrm{s})^{-1} \\
        \Gamma_{60P}^{\rm BB}&\approx(161\,\mu\mathrm{s})^{-1}.
    \end{split}
    \label{eq:gamma_BB}
    \end{align} 
    We neglect the possibility of blackbody-induced transitions between the states of the computational basis ($\ket{\uparrow}$, $\ket{h}$ and $\ket{\downarrow}$), which rate is small compared with the decay rate to other states in the Rydberg manifold.
\end{enumerate}

{\it Positional disorder -}
At time~$T=0$ the tweezer light is turned-off, which projects the trapped spatial wavefunction onto free space leading to a dispersive wavepacket. Although we do not include the subsequent motion or spin-motion coupling, that are expected to be small at the experimental energy and timescales, we model the initial wavefunction spread by including positional disorder. We assume the spatial probability distribution of each atom to be normally distributed around its target position with a standard deviation of~$\sigma_{xy}=0.1\,\mu\mathrm{m}$ (in-plane) and  $\sigma_{z}=1.0\,\mu\mathrm{m}$ (perpendicular to the plane). For each numerical trajectory, we sample the atom's initial positions and time evolve the internal states under the Rydberg Hamiltonian with couplings according to the atom's spatial distribution.
\\
\\
\noindent{\bf Numerical simulations: 2D AFM.}
To simulate the dynamics of the initial 2D AFM product state, we apply matrix-product-state (MPS) methods. In particular, we time evolve under Hamiltonian~\eqref{eq:bare-Ryd} using the built-in two-site time-dependent variation principle algorithm (2DTVP) of the TenPy package~\cite{Hauschild2018,Hauschild2019}. All calculations use a maximum bond dimension~$m=1000$ and we choose the time steps~$\Delta T = 0.025 \times \frac{2\pi}{|2t|}$. To account for experimental imperfections, we average the local hole probability, see Fig.\,\ref{fig:2d}c and d, over $400$ trajectories according to initial product states sampled from the the error budget shown in Fig.~\ref{fig:SM_error_2d}b. As an indicator for convergence, we analyze the energy density~$
e(T) = \bra{\psi(T)} \hat{H} \ket{\psi(T)}/L$ and we confirm that the deviation of the energy density to the initial state energy density~$e(0)$ remains within $\mathrm{max}_T |e(T)- e(0)| \leq 1\cdot 10^{-4} \cdot |t|$ ($\mathrm{max}_T |e(T)- e(0)| \leq 7\cdot 10^{-5} \cdot |t|$) for simulations with long-range Rydberg tails (only NN interactions).

\section*{Data Availability}

All data are available from the corresponding author on request.

\section*{Competing interests}
Antoine Browaeys and T.L. are cofounders and shareholders of PASQAL. The remaining authors declare no competing interests.

\subsection*{Acknowledgements}
We thank Johannes Mögerle, Katarina Brechtelsbauer, and Hans-Peter Büchler for insightful discussions about encoding spins using three Rydberg states, 
and Johannes Mögerle for help regarding the use of the Pairinteraction software package~\cite{Weber2017}.
This work is supported by the Agence Nationale de la Recherche (ANR-22-PETQ-0004 France 2030, project QuBitAF), the European Research Council (Advanced grant No. 101018511-ATARAXIA), and the Horizon Europe programme HORIZON-CL4- 2022-QUANTUM-02-SGA (project 101113690 (PASQuanS2.1). RM acknowledges funding by the “Fondation CFM pour la Recherche” via a Jean-Pierre Aguilar PhD scholarship. LH is supported by the Simons Collaboration on Ultra-Quantum Matter, which is a grant from the Simons Foundation (651440).
LH and FG were funded by the Deutsche Forschungsgemeinschaft (DFG, German Research Foundation)
under Germany’s Excellence Strategy – EXC-2111 –
390814868 and have received funding from the European
Research Council (ERC) under the European Union’s
Horizon 2020 research and innovation programm (Grant
Agreement no 948141) — ERC Starting Grant SimUcQuam. SH acknowledges funding through the Harvard Quantum Initiative Postdoctoral Fellowship in Quantum Science and Engineering. SG acknowledges funding by Structures (Deutsche Forschungsgemeinschaft (DFG, German Research Foundation) under Germany’s Excellence Strategy
EXC2181/1-390900948). DB acknowledges support from MCIN/AEI/10.13039/501100011033 (PID2020-119667GA-I00, CNS2022-135781, EUR2022-134067 and European Union NextGenerationEU PRTR-C17.I1).

\section*{Author contributions}
\noindent
$^*$ MQ, GE, CC, LH, SH contributed equally to this work.
MQ, GE, CC, GB, RM carried out the experiments with the help of BG, LK, DB. 
LH and MQ conducted the numerical simulations. 
LH, SH, SG, NC devised the Rydberg scheme combining dipolar and van-der-Waals couplings to realize the $t-J$ model. 
Annabelle Bohrdt, FG, TL, Antoine Browaeys supervised the work.
All authors contributed to the data analysis, progression of the project, 
and on both the experimental and theoretical side. 
All authors contributed to the writing of the manuscript. 
Correspondence and requests for materials should be addressed to Antoine Browaeys or Mu Qiao.


\begin{thebibliography}{47}%
\makeatletter
\providecommand \@ifxundefined [1]{%
 \@ifx{#1\undefined}
}%
\providecommand \@ifnum [1]{%
 \ifnum #1\expandafter \@firstoftwo
 \else \expandafter \@secondoftwo
 \fi
}%
\providecommand \@ifx [1]{%
 \ifx #1\expandafter \@firstoftwo
 \else \expandafter \@secondoftwo
 \fi
}%
\providecommand \natexlab [1]{#1}%
\providecommand \enquote  [1]{``#1''}%
\providecommand \bibnamefont  [1]{#1}%
\providecommand \bibfnamefont [1]{#1}%
\providecommand \citenamefont [1]{#1}%
\providecommand \href@noop [0]{\@secondoftwo}%
\providecommand \href [0]{\begingroup \@sanitize@url \@href}%
\providecommand \@href[1]{\@@startlink{#1}\@@href}%
\providecommand \@@href[1]{\endgroup#1\@@endlink}%
\providecommand \@sanitize@url [0]{\catcode `\\12\catcode `\$12\catcode `\&12\catcode `\#12\catcode `\^12\catcode `\_12\catcode `\%12\relax}%
\providecommand \@@startlink[1]{}%
\providecommand \@@endlink[0]{}%
\providecommand \url  [0]{\begingroup\@sanitize@url \@url }%
\providecommand \@url [1]{\endgroup\@href {#1}{\urlprefix }}%
\providecommand \urlprefix  [0]{URL }%
\providecommand \Eprint [0]{\href }%
\providecommand \doibase [0]{https://doi.org/}%
\providecommand \selectlanguage [0]{\@gobble}%
\providecommand \bibinfo  [0]{\@secondoftwo}%
\providecommand \bibfield  [0]{\@secondoftwo}%
\providecommand \translation [1]{[#1]}%
\providecommand \BibitemOpen [0]{}%
\providecommand \bibitemStop [0]{}%
\providecommand \bibitemNoStop [0]{.\EOS\space}%
\providecommand \EOS [0]{\spacefactor3000\relax}%
\providecommand \BibitemShut  [1]{\csname bibitem#1\endcsname}%
\let\auto@bib@innerbib\@empty
\bibitem [{\citenamefont {Lee}\ \emph {et~al.}(2006)\citenamefont {Lee}, \citenamefont {Nagaosa},\ and\ \citenamefont {Wen}}]{Lee2006}%
  \BibitemOpen
  \bibfield  {author} {\bibinfo {author} {\bibfnamefont {P.~A.}\ \bibnamefont {Lee}}, \bibinfo {author} {\bibfnamefont {N.}~\bibnamefont {Nagaosa}},\ and\ \bibinfo {author} {\bibfnamefont {X.-G.}\ \bibnamefont {Wen}},\ }\bibfield  {title} {\bibinfo {title} {Doping a {M}ott insulator: Physics of high-temperature superconductivity},\ }\href {https://link.aps.org/doi/10.1103/RevModPhys.78.17} {\bibfield  {journal} {\bibinfo  {journal} {Rev. Mod. Phys.}\ }\textbf {\bibinfo {volume} {78}},\ \bibinfo {pages} {17} (\bibinfo {year} {2006})}\BibitemShut {NoStop}%
\bibitem [{\citenamefont {Bohrdt}\ \emph {et~al.}(2021)\citenamefont {Bohrdt}, \citenamefont {Homeier}, \citenamefont {Reinmoser}, \citenamefont {Demler},\ and\ \citenamefont {Grusdt}}]{Bohrdt2021}%
  \BibitemOpen
  \bibfield  {author} {\bibinfo {author} {\bibfnamefont {A.}~\bibnamefont {Bohrdt}}, \bibinfo {author} {\bibfnamefont {L.}~\bibnamefont {Homeier}}, \bibinfo {author} {\bibfnamefont {C.}~\bibnamefont {Reinmoser}}, \bibinfo {author} {\bibfnamefont {E.}~\bibnamefont {Demler}},\ and\ \bibinfo {author} {\bibfnamefont {F.}~\bibnamefont {Grusdt}},\ }\bibfield  {title} {\bibinfo {title} {Exploration of doped quantum magnets with ultracold atoms},\ }\href {https://doi.org/10.1016/j.aop.2021.168651} {\bibfield  {journal} {\bibinfo  {journal} {Annals of Physics}\ }\textbf {\bibinfo {volume} {435}},\ \bibinfo {pages} {168651} (\bibinfo {year} {2021})}\BibitemShut {NoStop}%
\bibitem [{\citenamefont {Auerbach}(1994)}]{Auerbach1994}%
  \BibitemOpen
  \bibfield  {author} {\bibinfo {author} {\bibfnamefont {A.}~\bibnamefont {Auerbach}},\ }\href {https://doi.org/10.1007/978-1-4612-0869-3} {\emph {\bibinfo {title} {Interacting {E}lectrons and {Q}uantum {M}agnetism}}}\ (\bibinfo  {publisher} {Springer New York},\ \bibinfo {year} {1994})\BibitemShut {NoStop}%
\bibitem [{\citenamefont {Carroll}\ \emph {et~al.}(2025)\citenamefont {Carroll}, \citenamefont {Hirzler}, \citenamefont {Miller}, \citenamefont {Wellnitz}, \citenamefont {Muleady}, \citenamefont {Lin}, \citenamefont {Zamarski}, \citenamefont {Wang}, \citenamefont {Bohn}, \citenamefont {Rey},\ and\ \citenamefont {Ye}}]{junye_tJ}%
  \BibitemOpen
  \bibfield  {author} {\bibinfo {author} {\bibfnamefont {A.~N.}\ \bibnamefont {Carroll}}, \bibinfo {author} {\bibfnamefont {H.}~\bibnamefont {Hirzler}}, \bibinfo {author} {\bibfnamefont {C.}~\bibnamefont {Miller}}, \bibinfo {author} {\bibfnamefont {D.}~\bibnamefont {Wellnitz}}, \bibinfo {author} {\bibfnamefont {S.~R.}\ \bibnamefont {Muleady}}, \bibinfo {author} {\bibfnamefont {J.}~\bibnamefont {Lin}}, \bibinfo {author} {\bibfnamefont {K.~P.}\ \bibnamefont {Zamarski}}, \bibinfo {author} {\bibfnamefont {R.~R.~W.}\ \bibnamefont {Wang}}, \bibinfo {author} {\bibfnamefont {J.~L.}\ \bibnamefont {Bohn}}, \bibinfo {author} {\bibfnamefont {A.~M.}\ \bibnamefont {Rey}},\ and\ \bibinfo {author} {\bibfnamefont {J.}~\bibnamefont {Ye}},\ }\bibfield  {title} {\bibinfo {title} {{Observation of generalized t-J spin dynamics with tunable dipolar interactions}},\ }\href {https://doi.org/10.1126/science.adq0911} {\bibfield  {journal} {\bibinfo  {journal} {Science}\ }\textbf {\bibinfo {volume} {388}},\ \bibinfo {pages} {381}
  (\bibinfo {year} {2025})}\BibitemShut {NoStop}%
\bibitem [{\citenamefont {Gorshkov}\ \emph {et~al.}(2011)\citenamefont {Gorshkov}, \citenamefont {Manmana}, \citenamefont {Chen}, \citenamefont {Ye}, \citenamefont {Demler}, \citenamefont {Lukin},\ and\ \citenamefont {Rey}}]{Gorshkov_2011_tJVW}%
  \BibitemOpen
  \bibfield  {author} {\bibinfo {author} {\bibfnamefont {A.~V.}\ \bibnamefont {Gorshkov}}, \bibinfo {author} {\bibfnamefont {S.~R.}\ \bibnamefont {Manmana}}, \bibinfo {author} {\bibfnamefont {G.}~\bibnamefont {Chen}}, \bibinfo {author} {\bibfnamefont {J.}~\bibnamefont {Ye}}, \bibinfo {author} {\bibfnamefont {E.}~\bibnamefont {Demler}}, \bibinfo {author} {\bibfnamefont {M.~D.}\ \bibnamefont {Lukin}},\ and\ \bibinfo {author} {\bibfnamefont {A.~M.}\ \bibnamefont {Rey}},\ }\bibfield  {title} {\bibinfo {title} {Tunable superfluidity and quantum magnetism with ultracold polar molecules},\ }\href {https://doi.org/10.1103/PhysRevLett.107.115301} {\bibfield  {journal} {\bibinfo  {journal} {Phys. Rev. Lett.}\ }\textbf {\bibinfo {volume} {107}},\ \bibinfo {pages} {115301} (\bibinfo {year} {2011})}\BibitemShut {NoStop}%
\bibitem [{\citenamefont {Qin}\ \emph {et~al.}(2020)\citenamefont {Qin}, \citenamefont {Chung}, \citenamefont {Shi}, \citenamefont {Vitali}, \citenamefont {Hubig}, \citenamefont {Schollw\"ock}, \citenamefont {White},\ and\ \citenamefont {Zhang}}]{qin2020absence}%
  \BibitemOpen
  \bibfield  {author} {\bibinfo {author} {\bibfnamefont {M.}~\bibnamefont {Qin}}, \bibinfo {author} {\bibfnamefont {C.-M.}\ \bibnamefont {Chung}}, \bibinfo {author} {\bibfnamefont {H.}~\bibnamefont {Shi}}, \bibinfo {author} {\bibfnamefont {E.}~\bibnamefont {Vitali}}, \bibinfo {author} {\bibfnamefont {C.}~\bibnamefont {Hubig}}, \bibinfo {author} {\bibfnamefont {U.}~\bibnamefont {Schollw\"ock}}, \bibinfo {author} {\bibfnamefont {S.~R.}\ \bibnamefont {White}},\ and\ \bibinfo {author} {\bibfnamefont {S.}~\bibnamefont {Zhang}} (\bibinfo {collaboration} {Simons Collaboration on the Many-Electron Problem}),\ }\bibfield  {title} {\bibinfo {title} {Absence of superconductivity in the pure two-dimensional {Hubbard} model},\ }\href {https://doi.org/10.1103/PhysRevX.10.031016} {\bibfield  {journal} {\bibinfo  {journal} {Phys. Rev. X}\ }\textbf {\bibinfo {volume} {10}},\ \bibinfo {pages} {031016} (\bibinfo {year} {2020})}\BibitemShut {NoStop}%
\bibitem [{\citenamefont {Jiang}\ and\ \citenamefont {Devereaux}(2019)}]{jiang_t'_2019}%
  \BibitemOpen
  \bibfield  {author} {\bibinfo {author} {\bibfnamefont {H.-C.}\ \bibnamefont {Jiang}}\ and\ \bibinfo {author} {\bibfnamefont {T.~P.}\ \bibnamefont {Devereaux}},\ }\bibfield  {title} {\bibinfo {title} {Superconductivity in the doped {H}ubbard model and its interplay with next-nearest hopping t'},\ }\href {https://doi.org/10.1126/science.aal5304} {\bibfield  {journal} {\bibinfo  {journal} {Science}\ }\textbf {\bibinfo {volume} {365}},\ \bibinfo {pages} {1424} (\bibinfo {year} {2019})}\BibitemShut {NoStop}%
\bibitem [{\citenamefont {Jiang}\ \emph {et~al.}(2024)\citenamefont {Jiang}, \citenamefont {Devereaux},\ and\ \citenamefont {Jiang}}]{jiang_t'_2024}%
  \BibitemOpen
  \bibfield  {author} {\bibinfo {author} {\bibfnamefont {Y.-F.}\ \bibnamefont {Jiang}}, \bibinfo {author} {\bibfnamefont {T.~P.}\ \bibnamefont {Devereaux}},\ and\ \bibinfo {author} {\bibfnamefont {H.-C.}\ \bibnamefont {Jiang}},\ }\bibfield  {title} {\bibinfo {title} {Ground-state phase diagram and superconductivity of the doped {H}ubbard model on six-leg square cylinders},\ }\href {https://doi.org/10.1103/PhysRevB.109.085121} {\bibfield  {journal} {\bibinfo  {journal} {Phys. Rev. B}\ }\textbf {\bibinfo {volume} {109}},\ \bibinfo {pages} {085121} (\bibinfo {year} {2024})}\BibitemShut {NoStop}%
\bibitem [{\citenamefont {Xu}\ \emph {et~al.}(2024)\citenamefont {Xu}, \citenamefont {Chung}, \citenamefont {Qin}, \citenamefont {Schollwöck}, \citenamefont {White},\ and\ \citenamefont {Zhang}}]{Xu2024}%
  \BibitemOpen
  \bibfield  {author} {\bibinfo {author} {\bibfnamefont {H.}~\bibnamefont {Xu}}, \bibinfo {author} {\bibfnamefont {C.-M.}\ \bibnamefont {Chung}}, \bibinfo {author} {\bibfnamefont {M.}~\bibnamefont {Qin}}, \bibinfo {author} {\bibfnamefont {U.}~\bibnamefont {Schollwöck}}, \bibinfo {author} {\bibfnamefont {S.~R.}\ \bibnamefont {White}},\ and\ \bibinfo {author} {\bibfnamefont {S.}~\bibnamefont {Zhang}},\ }\bibfield  {title} {\bibinfo {title} {{Coexistence of superconductivity with partially filled stripes in the Hubbard model}},\ }\href {https://doi.org/10.1126/science.adh7691} {\bibfield  {journal} {\bibinfo  {journal} {Science}\ }\textbf {\bibinfo {volume} {384}} (\bibinfo {year} {2024})}\BibitemShut {NoStop}%
\bibitem [{\citenamefont {Bespalova}\ \emph {et~al.}(2024)\citenamefont {Bespalova}, \citenamefont {Delić}, \citenamefont {Pupillo}, \citenamefont {Tacchino},\ and\ \citenamefont {Tavernelli}}]{Bespalova2024}%
  \BibitemOpen
  \bibfield  {author} {\bibinfo {author} {\bibfnamefont {T.~A.}\ \bibnamefont {Bespalova}}, \bibinfo {author} {\bibfnamefont {K.}~\bibnamefont {Delić}}, \bibinfo {author} {\bibfnamefont {G.}~\bibnamefont {Pupillo}}, \bibinfo {author} {\bibfnamefont {F.}~\bibnamefont {Tacchino}},\ and\ \bibinfo {author} {\bibfnamefont {I.}~\bibnamefont {Tavernelli}},\ }\href {https://arxiv.org/abs/2410.07789} {\bibinfo {title} {Simulating the {F}ermi-{H}ubbard model with long-range hopping on a quantum computer}} (\bibinfo {year} {2024}),\ \Eprint {https://arxiv.org/abs/2410.07789} {arXiv:2410.07789} \BibitemShut {NoStop}%
\bibitem [{\citenamefont {Bohrdt}\ \emph {et~al.}(2024)\citenamefont {Bohrdt}, \citenamefont {Wei}, \citenamefont {Adler}, \citenamefont {Srakaew}, \citenamefont {Agrawal}, \citenamefont {Weckesser}, \citenamefont {Bloch}, \citenamefont {Grusdt},\ and\ \citenamefont {Zeiher}}]{Bohrdt2024}%
  \BibitemOpen
  \bibfield  {author} {\bibinfo {author} {\bibfnamefont {A.}~\bibnamefont {Bohrdt}}, \bibinfo {author} {\bibfnamefont {D.}~\bibnamefont {Wei}}, \bibinfo {author} {\bibfnamefont {D.}~\bibnamefont {Adler}}, \bibinfo {author} {\bibfnamefont {K.}~\bibnamefont {Srakaew}}, \bibinfo {author} {\bibfnamefont {S.}~\bibnamefont {Agrawal}}, \bibinfo {author} {\bibfnamefont {P.}~\bibnamefont {Weckesser}}, \bibinfo {author} {\bibfnamefont {I.}~\bibnamefont {Bloch}}, \bibinfo {author} {\bibfnamefont {F.}~\bibnamefont {Grusdt}},\ and\ \bibinfo {author} {\bibfnamefont {J.}~\bibnamefont {Zeiher}},\ }\href {https://arxiv.org/abs/2410.19500} {\bibinfo {title} {{Microscopy of bosonic charge carriers in staggered magnetic fields}}} (\bibinfo {year} {2024}),\ \Eprint {https://arxiv.org/abs/2410.19500} {arXiv:2410.19500} \BibitemShut {NoStop}%
\bibitem [{\citenamefont {Homeier}\ \emph {et~al.}(2024)\citenamefont {Homeier}, \citenamefont {Harris}, \citenamefont {Blatz}, \citenamefont {Geier}, \citenamefont {Hollerith}, \citenamefont {Schollw\"ock}, \citenamefont {Grusdt},\ and\ \citenamefont {Bohrdt}}]{Lukas_2024_tJ}%
  \BibitemOpen
  \bibfield  {author} {\bibinfo {author} {\bibfnamefont {L.}~\bibnamefont {Homeier}}, \bibinfo {author} {\bibfnamefont {T.~J.}\ \bibnamefont {Harris}}, \bibinfo {author} {\bibfnamefont {T.}~\bibnamefont {Blatz}}, \bibinfo {author} {\bibfnamefont {S.}~\bibnamefont {Geier}}, \bibinfo {author} {\bibfnamefont {S.}~\bibnamefont {Hollerith}}, \bibinfo {author} {\bibfnamefont {U.}~\bibnamefont {Schollw\"ock}}, \bibinfo {author} {\bibfnamefont {F.}~\bibnamefont {Grusdt}},\ and\ \bibinfo {author} {\bibfnamefont {A.}~\bibnamefont {Bohrdt}},\ }\bibfield  {title} {\bibinfo {title} {Antiferromagnetic bosonic $\mathit{t}\ensuremath{-}\mathit{J}$ models and their quantum simulation in tweezer arrays},\ }\href {https://doi.org/10.1103/PhysRevLett.132.230401} {\bibfield  {journal} {\bibinfo  {journal} {Phys. Rev. Lett.}\ }\textbf {\bibinfo {volume} {132}},\ \bibinfo {pages} {230401} (\bibinfo {year} {2024})}\BibitemShut {NoStop}%
\bibitem [{\citenamefont {Browaeys}\ and\ \citenamefont {Lahaye}(2020)}]{Browaeys2020_NP}%
  \BibitemOpen
  \bibfield  {author} {\bibinfo {author} {\bibfnamefont {A.}~\bibnamefont {Browaeys}}\ and\ \bibinfo {author} {\bibfnamefont {T.}~\bibnamefont {Lahaye}},\ }\bibfield  {title} {\bibinfo {title} {Many-body physics with individually controlled {R}ydberg atoms},\ }\href {https://doi.org/10.1038/s41567-019-0733-z} {\bibfield  {journal} {\bibinfo  {journal} {Nature Physics}\ }\textbf {\bibinfo {volume} {16}},\ \bibinfo {pages} {132} (\bibinfo {year} {2020})}\BibitemShut {NoStop}%
\bibitem [{\citenamefont {M\"ogerle}\ \emph {et~al.}(2025)\citenamefont {M\"ogerle}, \citenamefont {Brechtelsbauer}, \citenamefont {Gea-Caballero}, \citenamefont {Prior}, \citenamefont {Emperauger}, \citenamefont {Bornet}, \citenamefont {Chen}, \citenamefont {Lahaye}, \citenamefont {Browaeys},\ and\ \citenamefont {B\"uchler}}]{Mogerle2024}%
  \BibitemOpen
  \bibfield  {author} {\bibinfo {author} {\bibfnamefont {J.}~\bibnamefont {M\"ogerle}}, \bibinfo {author} {\bibfnamefont {K.}~\bibnamefont {Brechtelsbauer}}, \bibinfo {author} {\bibfnamefont {A.}~\bibnamefont {Gea-Caballero}}, \bibinfo {author} {\bibfnamefont {J.}~\bibnamefont {Prior}}, \bibinfo {author} {\bibfnamefont {G.}~\bibnamefont {Emperauger}}, \bibinfo {author} {\bibfnamefont {G.}~\bibnamefont {Bornet}}, \bibinfo {author} {\bibfnamefont {C.}~\bibnamefont {Chen}}, \bibinfo {author} {\bibfnamefont {T.}~\bibnamefont {Lahaye}}, \bibinfo {author} {\bibfnamefont {A.}~\bibnamefont {Browaeys}},\ and\ \bibinfo {author} {\bibfnamefont {H.}~\bibnamefont {B\"uchler}},\ }\bibfield  {title} {\bibinfo {title} {Spin-1 haldane phase in a chain of rydberg atoms},\ }\href {https://doi.org/10.1103/PRXQuantum.6.020332} {\bibfield  {journal} {\bibinfo  {journal} {PRX Quantum}\ }\textbf {\bibinfo {volume} {6}},\ \bibinfo {pages} {020332} (\bibinfo {year} {2025})}\BibitemShut {NoStop}%
\bibitem [{\citenamefont {Liu}\ \emph {et~al.}(2024)\citenamefont {Liu}, \citenamefont {Bintz}, \citenamefont {Block}, \citenamefont {Samajdar}, \citenamefont {Kemp},\ and\ \citenamefont {Yao}}]{Liu2024}%
  \BibitemOpen
  \bibfield  {author} {\bibinfo {author} {\bibfnamefont {V.~S.}\ \bibnamefont {Liu}}, \bibinfo {author} {\bibfnamefont {M.}~\bibnamefont {Bintz}}, \bibinfo {author} {\bibfnamefont {M.}~\bibnamefont {Block}}, \bibinfo {author} {\bibfnamefont {R.}~\bibnamefont {Samajdar}}, \bibinfo {author} {\bibfnamefont {J.}~\bibnamefont {Kemp}},\ and\ \bibinfo {author} {\bibfnamefont {N.~Y.}\ \bibnamefont {Yao}},\ }\href {https://arxiv.org/abs/2407.17554} {\bibinfo {title} {{Supersolidity and Simplex Phases in Spin-1 Rydberg Atom Arrays}}} (\bibinfo {year} {2024}),\ \Eprint {https://arxiv.org/abs/2407.17554} {arXiv:2407.17554} \BibitemShut {NoStop}%
\bibitem [{\citenamefont {Scalapino}(2012)}]{Scalapino2012}%
  \BibitemOpen
  \bibfield  {author} {\bibinfo {author} {\bibfnamefont {D.~J.}\ \bibnamefont {Scalapino}},\ }\bibfield  {title} {\bibinfo {title} {A common thread: The pairing interaction for unconventional superconductors},\ }\href {https://doi.org/10.1103/revmodphys.84.1383} {\bibfield  {journal} {\bibinfo  {journal} {Reviews of Modern Physics}\ }\textbf {\bibinfo {volume} {84}},\ \bibinfo {pages} {1383} (\bibinfo {year} {2012})}\BibitemShut {NoStop}%
\bibitem [{\citenamefont {Proust}\ and\ \citenamefont {Taillefer}(2019)}]{Proust2019}%
  \BibitemOpen
  \bibfield  {author} {\bibinfo {author} {\bibfnamefont {C.}~\bibnamefont {Proust}}\ and\ \bibinfo {author} {\bibfnamefont {L.}~\bibnamefont {Taillefer}},\ }\bibfield  {title} {\bibinfo {title} {{The Remarkable Underlying Ground States of Cuprate Superconductors}},\ }\href {https://doi.org/10.1146/annurev-conmatphys-031218-013210} {\bibfield  {journal} {\bibinfo  {journal} {Annual Review of Condensed Matter Physics}\ }\textbf {\bibinfo {volume} {10}},\ \bibinfo {pages} {409} (\bibinfo {year} {2019})}\BibitemShut {NoStop}%
\bibitem [{\citenamefont {Bednorz}\ and\ \citenamefont {M{\"u}ller}(1986)}]{Bednorz1986}%
  \BibitemOpen
  \bibfield  {author} {\bibinfo {author} {\bibfnamefont {J.~G.}\ \bibnamefont {Bednorz}}\ and\ \bibinfo {author} {\bibfnamefont {K.~A.}\ \bibnamefont {M{\"u}ller}},\ }\bibfield  {title} {\bibinfo {title} {Possible high{T}c superconductivity in the {Ba-La-Cu-O} system},\ }\href {https://doi.org/10.1007/BF01303701} {\bibfield  {journal} {\bibinfo  {journal} {Zeitschrift f{\"u}r Physik B Condensed Matter}\ }\textbf {\bibinfo {volume} {64}},\ \bibinfo {pages} {189} (\bibinfo {year} {1986})}\BibitemShut {NoStop}%
\bibitem [{\citenamefont {Keimer}\ \emph {et~al.}(2015)\citenamefont {Keimer}, \citenamefont {Kivelson}, \citenamefont {Norman}, \citenamefont {Uchida},\ and\ \citenamefont {Zaanen}}]{Keimer2015}%
  \BibitemOpen
  \bibfield  {author} {\bibinfo {author} {\bibfnamefont {B.}~\bibnamefont {Keimer}}, \bibinfo {author} {\bibfnamefont {S.~A.}\ \bibnamefont {Kivelson}}, \bibinfo {author} {\bibfnamefont {M.~R.}\ \bibnamefont {Norman}}, \bibinfo {author} {\bibfnamefont {S.}~\bibnamefont {Uchida}},\ and\ \bibinfo {author} {\bibfnamefont {J.}~\bibnamefont {Zaanen}},\ }\bibfield  {title} {\bibinfo {title} {From quantum matter to high-temperature superconductivity in copper oxides},\ }\href {https://doi.org/10.1038/nature14165} {\bibfield  {journal} {\bibinfo  {journal} {Nature}\ }\textbf {\bibinfo {volume} {518}},\ \bibinfo {pages} {179} (\bibinfo {year} {2015})}\BibitemShut {NoStop}%
\bibitem [{\citenamefont {Bloch}\ \emph {et~al.}(2008)\citenamefont {Bloch}, \citenamefont {Dalibard},\ and\ \citenamefont {Zwerger}}]{Bloch2008_review}%
  \BibitemOpen
  \bibfield  {author} {\bibinfo {author} {\bibfnamefont {I.}~\bibnamefont {Bloch}}, \bibinfo {author} {\bibfnamefont {J.}~\bibnamefont {Dalibard}},\ and\ \bibinfo {author} {\bibfnamefont {W.}~\bibnamefont {Zwerger}},\ }\bibfield  {title} {\bibinfo {title} {Many-body physics with ultracold gases},\ }\href {https://doi.org./10.1103/revmodphys.80.885} {\bibfield  {journal} {\bibinfo  {journal} {Reviews of Modern Physics}\ }\textbf {\bibinfo {volume} {80}},\ \bibinfo {pages} {885} (\bibinfo {year} {2008})}\BibitemShut {NoStop}%
\bibitem [{\citenamefont {Mazurenko}\ \emph {et~al.}(2017)\citenamefont {Mazurenko}, \citenamefont {Chiu}, \citenamefont {Ji}, \citenamefont {Parsons}, \citenamefont {Kan{\'a}sz-Nagy}, \citenamefont {Schmidt}, \citenamefont {Grusdt}, \citenamefont {Demler}, \citenamefont {Greif},\ and\ \citenamefont {Greiner}}]{Mazurenko2017}%
  \BibitemOpen
  \bibfield  {author} {\bibinfo {author} {\bibfnamefont {A.}~\bibnamefont {Mazurenko}}, \bibinfo {author} {\bibfnamefont {C.~S.}\ \bibnamefont {Chiu}}, \bibinfo {author} {\bibfnamefont {G.}~\bibnamefont {Ji}}, \bibinfo {author} {\bibfnamefont {M.~F.}\ \bibnamefont {Parsons}}, \bibinfo {author} {\bibfnamefont {M.}~\bibnamefont {Kan{\'a}sz-Nagy}}, \bibinfo {author} {\bibfnamefont {R.}~\bibnamefont {Schmidt}}, \bibinfo {author} {\bibfnamefont {F.}~\bibnamefont {Grusdt}}, \bibinfo {author} {\bibfnamefont {E.}~\bibnamefont {Demler}}, \bibinfo {author} {\bibfnamefont {D.}~\bibnamefont {Greif}},\ and\ \bibinfo {author} {\bibfnamefont {M.}~\bibnamefont {Greiner}},\ }\bibfield  {title} {\bibinfo {title} {A cold-atom {Fermi}--{Hubbard} antiferromagnet},\ }\href {https://doi.org/10.1038/nature22362} {\bibfield  {journal} {\bibinfo  {journal} {Nature}\ }\textbf {\bibinfo {volume} {545}},\ \bibinfo {pages} {462} (\bibinfo {year} {2017})}\BibitemShut {NoStop}%
\bibitem [{\citenamefont {Shao}\ \emph {et~al.}(2024)\citenamefont {Shao}, \citenamefont {Wang}, \citenamefont {Zhu}, \citenamefont {Zhu}, \citenamefont {Sun}, \citenamefont {Chen}, \citenamefont {Zhang}, \citenamefont {Fan}, \citenamefont {Deng}, \citenamefont {Yao}, \citenamefont {Chen},\ and\ \citenamefont {Pan}}]{Shao2024}%
  \BibitemOpen
  \bibfield  {author} {\bibinfo {author} {\bibfnamefont {H.-J.}\ \bibnamefont {Shao}}, \bibinfo {author} {\bibfnamefont {Y.-X.}\ \bibnamefont {Wang}}, \bibinfo {author} {\bibfnamefont {D.-Z.}\ \bibnamefont {Zhu}}, \bibinfo {author} {\bibfnamefont {Y.-S.}\ \bibnamefont {Zhu}}, \bibinfo {author} {\bibfnamefont {H.-N.}\ \bibnamefont {Sun}}, \bibinfo {author} {\bibfnamefont {S.-Y.}\ \bibnamefont {Chen}}, \bibinfo {author} {\bibfnamefont {C.}~\bibnamefont {Zhang}}, \bibinfo {author} {\bibfnamefont {Z.-J.}\ \bibnamefont {Fan}}, \bibinfo {author} {\bibfnamefont {Y.}~\bibnamefont {Deng}}, \bibinfo {author} {\bibfnamefont {X.-C.}\ \bibnamefont {Yao}}, \bibinfo {author} {\bibfnamefont {Y.-A.}\ \bibnamefont {Chen}},\ and\ \bibinfo {author} {\bibfnamefont {J.-W.}\ \bibnamefont {Pan}},\ }\bibfield  {title} {\bibinfo {title} {Antiferromagnetic phase transition in a 3d fermionic {H}ubbard model},\ }\href {https://doi.org/10.1038/s41586-024-07689-2} {\bibfield  {journal} {\bibinfo  {journal} {Nature}\ }\textbf {\bibinfo
  {volume} {632}},\ \bibinfo {pages} {267} (\bibinfo {year} {2024})}\BibitemShut {NoStop}%
\bibitem [{\citenamefont {Hirthe}\ \emph {et~al.}(2023)\citenamefont {Hirthe}, \citenamefont {Chalopin}, \citenamefont {Bourgund}, \citenamefont {Bojovi{\'{c}}}, \citenamefont {Bohrdt}, \citenamefont {Demler}, \citenamefont {Grusdt}, \citenamefont {Bloch},\ and\ \citenamefont {Hilker}}]{Hirthe2023}%
  \BibitemOpen
  \bibfield  {author} {\bibinfo {author} {\bibfnamefont {S.}~\bibnamefont {Hirthe}}, \bibinfo {author} {\bibfnamefont {T.}~\bibnamefont {Chalopin}}, \bibinfo {author} {\bibfnamefont {D.}~\bibnamefont {Bourgund}}, \bibinfo {author} {\bibfnamefont {P.}~\bibnamefont {Bojovi{\'{c}}}}, \bibinfo {author} {\bibfnamefont {A.}~\bibnamefont {Bohrdt}}, \bibinfo {author} {\bibfnamefont {E.}~\bibnamefont {Demler}}, \bibinfo {author} {\bibfnamefont {F.}~\bibnamefont {Grusdt}}, \bibinfo {author} {\bibfnamefont {I.}~\bibnamefont {Bloch}},\ and\ \bibinfo {author} {\bibfnamefont {T.~A.}\ \bibnamefont {Hilker}},\ }\bibfield  {title} {\bibinfo {title} {Magnetically mediated hole pairing in fermionic ladders of ultracold atoms},\ }\href {https://doi.org/10.1038/s41586-022-05437-y} {\bibfield  {journal} {\bibinfo  {journal} {Nature}\ }\textbf {\bibinfo {volume} {613}},\ \bibinfo {pages} {463} (\bibinfo {year} {2023})}\BibitemShut {NoStop}%
\bibitem [{\citenamefont {Lebrat}\ \emph {et~al.}(2024)\citenamefont {Lebrat}, \citenamefont {Xu}, \citenamefont {Kendrick}, \citenamefont {Kale}, \citenamefont {Gang}, \citenamefont {Seetharaman}, \citenamefont {Morera}, \citenamefont {Khatami}, \citenamefont {Demler},\ and\ \citenamefont {Greiner}}]{Lebrat2024}%
  \BibitemOpen
  \bibfield  {author} {\bibinfo {author} {\bibfnamefont {M.}~\bibnamefont {Lebrat}}, \bibinfo {author} {\bibfnamefont {M.}~\bibnamefont {Xu}}, \bibinfo {author} {\bibfnamefont {L.~H.}\ \bibnamefont {Kendrick}}, \bibinfo {author} {\bibfnamefont {A.}~\bibnamefont {Kale}}, \bibinfo {author} {\bibfnamefont {Y.}~\bibnamefont {Gang}}, \bibinfo {author} {\bibfnamefont {P.}~\bibnamefont {Seetharaman}}, \bibinfo {author} {\bibfnamefont {I.}~\bibnamefont {Morera}}, \bibinfo {author} {\bibfnamefont {E.}~\bibnamefont {Khatami}}, \bibinfo {author} {\bibfnamefont {E.}~\bibnamefont {Demler}},\ and\ \bibinfo {author} {\bibfnamefont {M.}~\bibnamefont {Greiner}},\ }\bibfield  {title} {\bibinfo {title} {{Observation of Nagaoka polarons in a Fermi–Hubbard quantum simulator}},\ }\href {https://doi.org/10.1038/s41586-024-07272-9} {\bibfield  {journal} {\bibinfo  {journal} {Nature}\ }\textbf {\bibinfo {volume} {629}},\ \bibinfo {pages} {317} (\bibinfo {year} {2024})}\BibitemShut {NoStop}%
\bibitem [{\citenamefont {Prichard}\ \emph {et~al.}(2024)\citenamefont {Prichard}, \citenamefont {Spar}, \citenamefont {Morera}, \citenamefont {Demler}, \citenamefont {Yan},\ and\ \citenamefont {Bakr}}]{Prichard2024}%
  \BibitemOpen
  \bibfield  {author} {\bibinfo {author} {\bibfnamefont {M.~L.}\ \bibnamefont {Prichard}}, \bibinfo {author} {\bibfnamefont {B.~M.}\ \bibnamefont {Spar}}, \bibinfo {author} {\bibfnamefont {I.}~\bibnamefont {Morera}}, \bibinfo {author} {\bibfnamefont {E.}~\bibnamefont {Demler}}, \bibinfo {author} {\bibfnamefont {Z.~Z.}\ \bibnamefont {Yan}},\ and\ \bibinfo {author} {\bibfnamefont {W.~S.}\ \bibnamefont {Bakr}},\ }\bibfield  {title} {\bibinfo {title} {{Directly imaging spin polarons in a kinetically frustrated Hubbard system}},\ }\href {https://doi.org/10.1038/s41586-024-07356-6} {\bibfield  {journal} {\bibinfo  {journal} {Nature}\ }\textbf {\bibinfo {volume} {629}},\ \bibinfo {pages} {323} (\bibinfo {year} {2024})}\BibitemShut {NoStop}%
\bibitem [{\citenamefont {Duan}\ \emph {et~al.}(2003)\citenamefont {Duan}, \citenamefont {Demler},\ and\ \citenamefont {Lukin}}]{Duan2003}%
  \BibitemOpen
  \bibfield  {author} {\bibinfo {author} {\bibfnamefont {L.-M.}\ \bibnamefont {Duan}}, \bibinfo {author} {\bibfnamefont {E.}~\bibnamefont {Demler}},\ and\ \bibinfo {author} {\bibfnamefont {M.~D.}\ \bibnamefont {Lukin}},\ }\bibfield  {title} {\bibinfo {title} {Controlling {S}pin {E}xchange {I}nteractions of {U}ltracold {A}toms in {O}ptical {L}attices},\ }\href {https://doi.org/10.1103/physrevlett.91.090402} {\bibfield  {journal} {\bibinfo  {journal} {Physical Review Letters}\ }\textbf {\bibinfo {volume} {91}},\ \bibinfo {pages} {090402} (\bibinfo {year} {2003})}\BibitemShut {NoStop}%
\bibitem [{\citenamefont {Gross}\ and\ \citenamefont {Bloch}(2017)}]{Gross2017}%
  \BibitemOpen
  \bibfield  {author} {\bibinfo {author} {\bibfnamefont {C.}~\bibnamefont {Gross}}\ and\ \bibinfo {author} {\bibfnamefont {I.}~\bibnamefont {Bloch}},\ }\bibfield  {title} {\bibinfo {title} {Quantum simulations with ultracold atoms in optical lattices},\ }\href {https://doi.org/10.1126/science.aal3837} {\bibfield  {journal} {\bibinfo  {journal} {Science}\ }\textbf {\bibinfo {volume} {357}},\ \bibinfo {pages} {995} (\bibinfo {year} {2017})}\BibitemShut {NoStop}%
\bibitem [{\citenamefont {Emery}\ \emph {et~al.}(1990)\citenamefont {Emery}, \citenamefont {Kivelson},\ and\ \citenamefont {Lin}}]{emery1990separation}%
  \BibitemOpen
  \bibfield  {author} {\bibinfo {author} {\bibfnamefont {V.~J.}\ \bibnamefont {Emery}}, \bibinfo {author} {\bibfnamefont {S.~A.}\ \bibnamefont {Kivelson}},\ and\ \bibinfo {author} {\bibfnamefont {H.~Q.}\ \bibnamefont {Lin}},\ }\bibfield  {title} {\bibinfo {title} {Phase separation in the t-j model},\ }\href {https://doi.org/10.1103/PhysRevLett.64.475} {\bibfield  {journal} {\bibinfo  {journal} {Phys. Rev. Lett.}\ }\textbf {\bibinfo {volume} {64}},\ \bibinfo {pages} {475} (\bibinfo {year} {1990})}\BibitemShut {NoStop}%
\bibitem [{\citenamefont {Boninsegni}(2001)}]{Boninsegni2001}%
  \BibitemOpen
  \bibfield  {author} {\bibinfo {author} {\bibfnamefont {M.}~\bibnamefont {Boninsegni}},\ }\bibfield  {title} {\bibinfo {title} {{Phase Separation in Mixtures of Hard Core Bosons}},\ }\href {https://doi.org/10.1103/physrevlett.87.087201} {\bibfield  {journal} {\bibinfo  {journal} {Physical Review Letters}\ }\textbf {\bibinfo {volume} {87}},\ \bibinfo {pages} {087201} (\bibinfo {year} {2001})}\BibitemShut {NoStop}%
\bibitem [{\citenamefont {Sun}\ \emph {et~al.}(2021)\citenamefont {Sun}, \citenamefont {Yang}, \citenamefont {Wang}, \citenamefont {Zhou}, \citenamefont {Su}, \citenamefont {Dai}, \citenamefont {Yuan},\ and\ \citenamefont {Pan}}]{Sun2021}%
  \BibitemOpen
  \bibfield  {author} {\bibinfo {author} {\bibfnamefont {H.}~\bibnamefont {Sun}}, \bibinfo {author} {\bibfnamefont {B.}~\bibnamefont {Yang}}, \bibinfo {author} {\bibfnamefont {H.-Y.}\ \bibnamefont {Wang}}, \bibinfo {author} {\bibfnamefont {Z.-Y.}\ \bibnamefont {Zhou}}, \bibinfo {author} {\bibfnamefont {G.-X.}\ \bibnamefont {Su}}, \bibinfo {author} {\bibfnamefont {H.-N.}\ \bibnamefont {Dai}}, \bibinfo {author} {\bibfnamefont {Z.-S.}\ \bibnamefont {Yuan}},\ and\ \bibinfo {author} {\bibfnamefont {J.-W.}\ \bibnamefont {Pan}},\ }\bibfield  {title} {\bibinfo {title} {Realization of a bosonic antiferromagnet},\ }\href {https://doi.org/10.1038/s41567-021-01277-1} {\bibfield  {journal} {\bibinfo  {journal} {Nature Physics}\ }\textbf {\bibinfo {volume} {17}},\ \bibinfo {pages} {990} (\bibinfo {year} {2021})}\BibitemShut {NoStop}%
\bibitem [{\citenamefont {Jepsen}\ \emph {et~al.}(2021)\citenamefont {Jepsen}, \citenamefont {Ho}, \citenamefont {Amato-Grill}, \citenamefont {Dimitrova}, \citenamefont {Demler},\ and\ \citenamefont {Ketterle}}]{Jepsen2021}%
  \BibitemOpen
  \bibfield  {author} {\bibinfo {author} {\bibfnamefont {P.~N.}\ \bibnamefont {Jepsen}}, \bibinfo {author} {\bibfnamefont {W.~W.}\ \bibnamefont {Ho}}, \bibinfo {author} {\bibfnamefont {J.}~\bibnamefont {Amato-Grill}}, \bibinfo {author} {\bibfnamefont {I.}~\bibnamefont {Dimitrova}}, \bibinfo {author} {\bibfnamefont {E.}~\bibnamefont {Demler}},\ and\ \bibinfo {author} {\bibfnamefont {W.}~\bibnamefont {Ketterle}},\ }\bibfield  {title} {\bibinfo {title} {Transverse {S}pin {D}ynamics in the {A}nisotropic {H}eisenberg {M}odel {R}ealized with {U}ltracold {A}toms},\ }\href {https://doi.org/10.1103/physrevx.11.041054} {\bibfield  {journal} {\bibinfo  {journal} {Physical Review X}\ }\textbf {\bibinfo {volume} {11}},\ \bibinfo {pages} {041054} (\bibinfo {year} {2021})}\BibitemShut {NoStop}%
\bibitem [{\citenamefont {Harris}\ \emph {et~al.}(2024)\citenamefont {Harris}, \citenamefont {Schollwöck}, \citenamefont {Bohrdt},\ and\ \citenamefont {Grusdt}}]{Harris2024}%
  \BibitemOpen
  \bibfield  {author} {\bibinfo {author} {\bibfnamefont {T.~J.}\ \bibnamefont {Harris}}, \bibinfo {author} {\bibfnamefont {U.}~\bibnamefont {Schollwöck}}, \bibinfo {author} {\bibfnamefont {A.}~\bibnamefont {Bohrdt}},\ and\ \bibinfo {author} {\bibfnamefont {F.}~\bibnamefont {Grusdt}},\ }\href {https://arxiv.org/abs/2410.00904} {\bibinfo {title} {{Kinetic magnetism and stripe order in the doped AFM bosonic ${t-J}$ model}}} (\bibinfo {year} {2024}),\ \Eprint {https://arxiv.org/abs/2410.00904} {arXiv:2410.00904} \BibitemShut {NoStop}%
\bibitem [{\citenamefont {Zhang}\ \emph {et~al.}(2024)\citenamefont {Zhang}, \citenamefont {Zhang}, \citenamefont {Xu},\ and\ \citenamefont {Weng}}]{Zhang2024}%
  \BibitemOpen
  \bibfield  {author} {\bibinfo {author} {\bibfnamefont {H.-K.}\ \bibnamefont {Zhang}}, \bibinfo {author} {\bibfnamefont {J.-X.}\ \bibnamefont {Zhang}}, \bibinfo {author} {\bibfnamefont {J.-S.}\ \bibnamefont {Xu}},\ and\ \bibinfo {author} {\bibfnamefont {Z.-Y.}\ \bibnamefont {Weng}},\ }\href {https://arxiv.org/abs/2409.15424} {\bibinfo {title} {{Quantum-interference-induced pairing in antiferromagnetic bosonic $t$-$J$ model}}} (\bibinfo {year} {2024}),\ \Eprint {https://arxiv.org/abs/2409.15424} {arXiv:2409.15424} \BibitemShut {NoStop}%
\bibitem [{\citenamefont {Siller}\ \emph {et~al.}(2001)\citenamefont {Siller}, \citenamefont {Troyer}, \citenamefont {Rice},\ and\ \citenamefont {White}}]{Siller2001}%
  \BibitemOpen
  \bibfield  {author} {\bibinfo {author} {\bibfnamefont {T.}~\bibnamefont {Siller}}, \bibinfo {author} {\bibfnamefont {M.}~\bibnamefont {Troyer}}, \bibinfo {author} {\bibfnamefont {T.~M.}\ \bibnamefont {Rice}},\ and\ \bibinfo {author} {\bibfnamefont {S.~R.}\ \bibnamefont {White}},\ }\bibfield  {title} {\bibinfo {title} {Bosonic model of hole pairs},\ }\href {https://doi.org/10.1103/physrevb.63.195106} {\bibfield  {journal} {\bibinfo  {journal} {Physical Review B}\ }\textbf {\bibinfo {volume} {63}},\ \bibinfo {pages} {195106} (\bibinfo {year} {2001})}\BibitemShut {NoStop}%
\bibitem [{\citenamefont {O'Mahony}\ \emph {et~al.}(2022)\citenamefont {O'Mahony}, \citenamefont {Ren}, \citenamefont {Chen}, \citenamefont {Chong}, \citenamefont {Liu}, \citenamefont {Eisaki}, \citenamefont {Uchida}, \citenamefont {Hamidian},\ and\ \citenamefont {Davis}}]{OMahony2022}%
  \BibitemOpen
  \bibfield  {author} {\bibinfo {author} {\bibfnamefont {S.~M.}\ \bibnamefont {O'Mahony}}, \bibinfo {author} {\bibfnamefont {W.}~\bibnamefont {Ren}}, \bibinfo {author} {\bibfnamefont {W.}~\bibnamefont {Chen}}, \bibinfo {author} {\bibfnamefont {Y.~X.}\ \bibnamefont {Chong}}, \bibinfo {author} {\bibfnamefont {X.}~\bibnamefont {Liu}}, \bibinfo {author} {\bibfnamefont {H.}~\bibnamefont {Eisaki}}, \bibinfo {author} {\bibfnamefont {S.}~\bibnamefont {Uchida}}, \bibinfo {author} {\bibfnamefont {M.~H.}\ \bibnamefont {Hamidian}},\ and\ \bibinfo {author} {\bibfnamefont {J.~C.~S.}\ \bibnamefont {Davis}},\ }\bibfield  {title} {\bibinfo {title} {On the electron pairing mechanism of copper-oxide high temperature superconductivity},\ }\href {https://doi.org/10.1073/pnas.2207449119} {\bibfield  {journal} {\bibinfo  {journal} {Proceedings of the National Academy of Sciences}\ }\textbf {\bibinfo {volume} {119}} (\bibinfo {year} {2022})}\BibitemShut {NoStop}%
\bibitem [{\citenamefont {Morera}\ \emph {et~al.}(2024)\citenamefont {Morera}, \citenamefont {Bohrdt}, \citenamefont {Ho},\ and\ \citenamefont {Demler}}]{Morera2024}%
  \BibitemOpen
  \bibfield  {author} {\bibinfo {author} {\bibfnamefont {I.}~\bibnamefont {Morera}}, \bibinfo {author} {\bibfnamefont {A.}~\bibnamefont {Bohrdt}}, \bibinfo {author} {\bibfnamefont {W.~W.}\ \bibnamefont {Ho}},\ and\ \bibinfo {author} {\bibfnamefont {E.}~\bibnamefont {Demler}},\ }\bibfield  {title} {\bibinfo {title} {Attraction from kinetic frustration in ladder systems},\ }\href {https://doi.org/10.1103/physrevresearch.6.023196} {\bibfield  {journal} {\bibinfo  {journal} {Physical Review Research}\ }\textbf {\bibinfo {volume} {6}},\ \bibinfo {pages} {023196} (\bibinfo {year} {2024})}\BibitemShut {NoStop}%
\bibitem [{\citenamefont {Sous}\ and\ \citenamefont {Pretko}(2020)}]{Sous2020}%
  \BibitemOpen
  \bibfield  {author} {\bibinfo {author} {\bibfnamefont {J.}~\bibnamefont {Sous}}\ and\ \bibinfo {author} {\bibfnamefont {M.}~\bibnamefont {Pretko}},\ }\bibfield  {title} {\bibinfo {title} {Fractons from polarons},\ }\bibfield  {journal} {\bibinfo  {journal} {Physical Review B}\ }\textbf {\bibinfo {volume} {102}},\ \href {https://doi.org/10.1103/physrevb.102.214437} {10.1103/physrevb.102.214437} (\bibinfo {year} {2020})\BibitemShut {NoStop}%
\bibitem [{\citenamefont {Barredo}\ \emph {et~al.}(2015)\citenamefont {Barredo}, \citenamefont {Labuhn}, \citenamefont {Ravets}, \citenamefont {Lahaye}, \citenamefont {Browaeys},\ and\ \citenamefont {Adams}}]{Barredo2015}%
  \BibitemOpen
  \bibfield  {author} {\bibinfo {author} {\bibfnamefont {D.}~\bibnamefont {Barredo}}, \bibinfo {author} {\bibfnamefont {H.}~\bibnamefont {Labuhn}}, \bibinfo {author} {\bibfnamefont {S.}~\bibnamefont {Ravets}}, \bibinfo {author} {\bibfnamefont {T.}~\bibnamefont {Lahaye}}, \bibinfo {author} {\bibfnamefont {A.}~\bibnamefont {Browaeys}},\ and\ \bibinfo {author} {\bibfnamefont {C.~S.}\ \bibnamefont {Adams}},\ }\bibfield  {title} {\bibinfo {title} {Coherent excitation transfer in a spin chain of three rydberg atoms},\ }\href {https://doi.org/10.1103/PhysRevLett.114.113002} {\bibfield  {journal} {\bibinfo  {journal} {Phys. Rev. Lett.}\ }\textbf {\bibinfo {volume} {114}},\ \bibinfo {pages} {113002} (\bibinfo {year} {2015})}\BibitemShut {NoStop}%
\bibitem [{\citenamefont {Emperauger}\ \emph {et~al.}(2025)\citenamefont {Emperauger}, \citenamefont {Qiao}, \citenamefont {Bornet}, \citenamefont {Chen}, \citenamefont {Martin}, \citenamefont {Chew}, \citenamefont {G\'ely}, \citenamefont {Klein}, \citenamefont {Barredo}, \citenamefont {Browaeys},\ and\ \citenamefont {Lahaye}}]{GEinprep}%
  \BibitemOpen
  \bibfield  {author} {\bibinfo {author} {\bibfnamefont {G.}~\bibnamefont {Emperauger}}, \bibinfo {author} {\bibfnamefont {M.}~\bibnamefont {Qiao}}, \bibinfo {author} {\bibfnamefont {G.}~\bibnamefont {Bornet}}, \bibinfo {author} {\bibfnamefont {C.}~\bibnamefont {Chen}}, \bibinfo {author} {\bibfnamefont {R.}~\bibnamefont {Martin}}, \bibinfo {author} {\bibfnamefont {Y.~T.}\ \bibnamefont {Chew}}, \bibinfo {author} {\bibfnamefont {B.}~\bibnamefont {G\'ely}}, \bibinfo {author} {\bibfnamefont {L.}~\bibnamefont {Klein}}, \bibinfo {author} {\bibfnamefont {D.}~\bibnamefont {Barredo}}, \bibinfo {author} {\bibfnamefont {A.}~\bibnamefont {Browaeys}},\ and\ \bibinfo {author} {\bibfnamefont {T.}~\bibnamefont {Lahaye}},\ }\bibfield  {title} {\bibinfo {title} {Benchmarking direct and indirect dipolar spin-exchange interactions between two rydberg atoms},\ }\href {https://doi.org/10.1103/PhysRevA.111.062806} {\bibfield  {journal} {\bibinfo  {journal} {Phys. Rev. A}\ }\textbf {\bibinfo {volume} {111}},\ \bibinfo {pages}
  {062806} (\bibinfo {year} {2025})}\BibitemShut {NoStop}%
\bibitem [{\citenamefont {Wadenpfuhl}\ and\ \citenamefont {Adams}(2024)}]{Wadenpfuhl2024}%
  \BibitemOpen
  \bibfield  {author} {\bibinfo {author} {\bibfnamefont {K.}~\bibnamefont {Wadenpfuhl}}\ and\ \bibinfo {author} {\bibfnamefont {C.~S.}\ \bibnamefont {Adams}},\ }\href {https://arxiv.org/abs/2412.14861} {\bibinfo {title} {{Unravelling the Structures in the van der Waals Interactions of Alkali Rydberg Atoms}}} (\bibinfo {year} {2024}),\ \Eprint {https://arxiv.org/abs/2412.14861} {arXiv:2412.14861} \BibitemShut {NoStop}%
\bibitem [{\citenamefont {Winkler}\ \emph {et~al.}(2006)\citenamefont {Winkler}, \citenamefont {Thalhammer}, \citenamefont {Lang}, \citenamefont {Grimm}, \citenamefont {Hecker~Denschlag}, \citenamefont {Daley}, \citenamefont {Kantian}, \citenamefont {B{\"u}chler},\ and\ \citenamefont {Zoller}}]{Winkler_2006_repulsive}%
  \BibitemOpen
  \bibfield  {author} {\bibinfo {author} {\bibfnamefont {K.}~\bibnamefont {Winkler}}, \bibinfo {author} {\bibfnamefont {G.}~\bibnamefont {Thalhammer}}, \bibinfo {author} {\bibfnamefont {F.}~\bibnamefont {Lang}}, \bibinfo {author} {\bibfnamefont {R.}~\bibnamefont {Grimm}}, \bibinfo {author} {\bibfnamefont {J.}~\bibnamefont {Hecker~Denschlag}}, \bibinfo {author} {\bibfnamefont {A.~J.}\ \bibnamefont {Daley}}, \bibinfo {author} {\bibfnamefont {A.}~\bibnamefont {Kantian}}, \bibinfo {author} {\bibfnamefont {H.~P.}\ \bibnamefont {B{\"u}chler}},\ and\ \bibinfo {author} {\bibfnamefont {P.}~\bibnamefont {Zoller}},\ }\bibfield  {title} {\bibinfo {title} {Repulsively bound atom pairs in an optical lattice},\ }\href {https://doi.org/10.1038/nature04918} {\bibfield  {journal} {\bibinfo  {journal} {Nature}\ }\textbf {\bibinfo {volume} {441}},\ \bibinfo {pages} {853} (\bibinfo {year} {2006})}\BibitemShut {NoStop}%
\bibitem [{\citenamefont {Staszewski}\ and\ \citenamefont {Wietek}(2025)}]{Staszewski2024}%
  \BibitemOpen
  \bibfield  {author} {\bibinfo {author} {\bibfnamefont {L.}~\bibnamefont {Staszewski}}\ and\ \bibinfo {author} {\bibfnamefont {A.}~\bibnamefont {Wietek}},\ }\bibfield  {title} {\bibinfo {title} {Quench dynamics of stripes and phase separation in the two-dimensional $t$-$j$ model},\ }\href {https://doi.org/10.1103/nfmp-32tt} {\bibfield  {journal} {\bibinfo  {journal} {Phys. Rev. B}\ ,\ } (\bibinfo {year} {2025})}\BibitemShut {NoStop}%
\bibitem [{\citenamefont {White}\ and\ \citenamefont {Scalapino}(2000)}]{white2000separation}%
  \BibitemOpen
  \bibfield  {author} {\bibinfo {author} {\bibfnamefont {S.~R.}\ \bibnamefont {White}}\ and\ \bibinfo {author} {\bibfnamefont {D.~J.}\ \bibnamefont {Scalapino}},\ }\bibfield  {title} {\bibinfo {title} {Phase separation and stripe formation in the two-dimensional $t\ensuremath{-}j$ model: A comparison of numerical results},\ }\href {https://doi.org/10.1103/PhysRevB.61.6320} {\bibfield  {journal} {\bibinfo  {journal} {Phys. Rev. B}\ }\textbf {\bibinfo {volume} {61}},\ \bibinfo {pages} {6320} (\bibinfo {year} {2000})}\BibitemShut {NoStop}%
\bibitem [{\citenamefont {Kagan}\ \emph {et~al.}(2021)\citenamefont {Kagan}, \citenamefont {Kugel},\ and\ \citenamefont {Rakhmanov}}]{Kagan2021}%
  \BibitemOpen
  \bibfield  {author} {\bibinfo {author} {\bibfnamefont {M.~Y.}\ \bibnamefont {Kagan}}, \bibinfo {author} {\bibfnamefont {K.~I.}\ \bibnamefont {Kugel}},\ and\ \bibinfo {author} {\bibfnamefont {A.~L.}\ \bibnamefont {Rakhmanov}},\ }\bibfield  {title} {\bibinfo {title} {Electronic phase separation: Recent progress in the old problem},\ }\href {https://doi.org/10.1016/j.physrep.2021.02.004} {\bibfield  {journal} {\bibinfo  {journal} {Physics Reports}\ }\textbf {\bibinfo {volume} {916}},\ \bibinfo {pages} {1} (\bibinfo {year} {2021})}\BibitemShut {NoStop}%
\bibitem [{\citenamefont {Ji}\ \emph {et~al.}(2021)\citenamefont {Ji}, \citenamefont {Xu}, \citenamefont {Kendrick}, \citenamefont {Chiu}, \citenamefont {Brüggenjürgen}, \citenamefont {Greif}, \citenamefont {Bohrdt}, \citenamefont {Grusdt}, \citenamefont {Demler}, \citenamefont {Lebrat},\ and\ \citenamefont {Greiner}}]{Ji2021}%
  \BibitemOpen
  \bibfield  {author} {\bibinfo {author} {\bibfnamefont {G.}~\bibnamefont {Ji}}, \bibinfo {author} {\bibfnamefont {M.}~\bibnamefont {Xu}}, \bibinfo {author} {\bibfnamefont {L.~H.}\ \bibnamefont {Kendrick}}, \bibinfo {author} {\bibfnamefont {C.~S.}\ \bibnamefont {Chiu}}, \bibinfo {author} {\bibfnamefont {J.~C.}\ \bibnamefont {Brüggenjürgen}}, \bibinfo {author} {\bibfnamefont {D.}~\bibnamefont {Greif}}, \bibinfo {author} {\bibfnamefont {A.}~\bibnamefont {Bohrdt}}, \bibinfo {author} {\bibfnamefont {F.}~\bibnamefont {Grusdt}}, \bibinfo {author} {\bibfnamefont {E.}~\bibnamefont {Demler}}, \bibinfo {author} {\bibfnamefont {M.}~\bibnamefont {Lebrat}},\ and\ \bibinfo {author} {\bibfnamefont {M.}~\bibnamefont {Greiner}},\ }\bibfield  {title} {\bibinfo {title} {Coupling a {M}obile {H}ole to an {A}ntiferromagnetic {S}pin {B}ackground: {T}ransient {D}ynamics of a {M}agnetic {P}olaron},\ }\href {https://doi.org/10.1103/physrevx.11.021022} {\bibfield  {journal} {\bibinfo  {journal} {Physical Review X}\ }\textbf {\bibinfo
  {volume} {11}} (\bibinfo {year} {2021})}\BibitemShut {NoStop}%
\bibitem [{\citenamefont {Chen}\ \emph {et~al.}(2023)\citenamefont {Chen}, \citenamefont {Bornet}, \citenamefont {Bintz}, \citenamefont {Emperauger}, \citenamefont {Leclerc}, \citenamefont {Liu}, \citenamefont {Scholl}, \citenamefont {Barredo}, \citenamefont {Hauschild}, \citenamefont {Chatterjee}, \citenamefont {Schuler}, \citenamefont {L{\"a}uchli}, \citenamefont {Zaletel}, \citenamefont {Lahaye}, \citenamefont {Yao},\ and\ \citenamefont {Browaeys}}]{Chen2023}%
  \BibitemOpen
  \bibfield  {author} {\bibinfo {author} {\bibfnamefont {C.}~\bibnamefont {Chen}}, \bibinfo {author} {\bibfnamefont {G.}~\bibnamefont {Bornet}}, \bibinfo {author} {\bibfnamefont {M.}~\bibnamefont {Bintz}}, \bibinfo {author} {\bibfnamefont {G.}~\bibnamefont {Emperauger}}, \bibinfo {author} {\bibfnamefont {L.}~\bibnamefont {Leclerc}}, \bibinfo {author} {\bibfnamefont {V.~S.}\ \bibnamefont {Liu}}, \bibinfo {author} {\bibfnamefont {P.}~\bibnamefont {Scholl}}, \bibinfo {author} {\bibfnamefont {D.}~\bibnamefont {Barredo}}, \bibinfo {author} {\bibfnamefont {J.}~\bibnamefont {Hauschild}}, \bibinfo {author} {\bibfnamefont {S.}~\bibnamefont {Chatterjee}}, \bibinfo {author} {\bibfnamefont {M.}~\bibnamefont {Schuler}}, \bibinfo {author} {\bibfnamefont {A.~M.}\ \bibnamefont {L{\"a}uchli}}, \bibinfo {author} {\bibfnamefont {M.~P.}\ \bibnamefont {Zaletel}}, \bibinfo {author} {\bibfnamefont {T.}~\bibnamefont {Lahaye}}, \bibinfo {author} {\bibfnamefont {N.~Y.}\ \bibnamefont {Yao}},\ and\ \bibinfo {author} {\bibfnamefont
  {A.}~\bibnamefont {Browaeys}},\ }\bibfield  {title} {\bibinfo {title} {Continuous symmetry breaking in a two-dimensional {R}ydberg array},\ }\href {https://doi.org/10.1038/s41586-023-05859-2} {\bibfield  {journal} {\bibinfo  {journal} {Nature}\ }\textbf {\bibinfo {volume} {616}},\ \bibinfo {pages} {691} (\bibinfo {year} {2023})}\BibitemShut {NoStop}%
\end{thebibliography}

\begin{thebibliography}{5}%
\makeatletter
\setcounter{NAT@ctr}{46}
\providecommand \@ifxundefined [1]{%
 \@ifx{#1\undefined}
}%
\providecommand \@ifnum [1]{%
 \ifnum #1\expandafter \@firstoftwo
 \else \expandafter \@secondoftwo
 \fi
}%
\providecommand \@ifx [1]{%
 \ifx #1\expandafter \@firstoftwo
 \else \expandafter \@secondoftwo
 \fi
}%
\providecommand \natexlab [1]{#1}%
\providecommand \enquote  [1]{``#1''}%
\providecommand \bibnamefont  [1]{#1}%
\providecommand \bibfnamefont [1]{#1}%
\providecommand \citenamefont [1]{#1}%
\providecommand \href@noop [0]{\@secondoftwo}%
\providecommand \href [0]{\begingroup \@sanitize@url \@href}%
\providecommand \@href[1]{\@@startlink{#1}\@@href}%
\providecommand \@@href[1]{\endgroup#1\@@endlink}%
\providecommand \@sanitize@url [0]{\catcode `\\12\catcode `\$12\catcode `\&12\catcode `\#12\catcode `\^12\catcode `\_12\catcode `\%12\relax}%
\providecommand \@@startlink[1]{}%
\providecommand \@@endlink[0]{}%
\providecommand \url  [0]{\begingroup\@sanitize@url \@url }%
\providecommand \@url [1]{\endgroup\@href {#1}{\urlprefix }}%
\providecommand \urlprefix  [0]{URL }%
\providecommand \Eprint [0]{\href }%
\providecommand \doibase [0]{https://doi.org/}%
\providecommand \selectlanguage [0]{\@gobble}%
\providecommand \bibinfo  [0]{\@secondoftwo}%
\providecommand \bibfield  [0]{\@secondoftwo}%
\providecommand \translation [1]{[#1]}%
\providecommand \BibitemOpen [0]{}%
\providecommand \bibitemStop [0]{}%
\providecommand \bibitemNoStop [0]{.\EOS\space}%
\providecommand \EOS [0]{\spacefactor3000\relax}%
\providecommand \BibitemShut  [1]{\csname bibitem#1\endcsname}%
\let\auto@bib@innerbib\@empty
\bibitem [{\citenamefont {Weber}\ \emph {et~al.}(2017)\citenamefont {Weber}, \citenamefont {Tresp}, \citenamefont {Menke}, \citenamefont {Urvoy}, \citenamefont {Firstenberg}, \citenamefont {B{\"u}chler},\ and\ \citenamefont {Hofferberth}}]{Weber2017}%
  \BibitemOpen
  \bibfield  {author} {\bibinfo {author} {\bibfnamefont {S.}~\bibnamefont {Weber}}, \bibinfo {author} {\bibfnamefont {C.}~\bibnamefont {Tresp}}, \bibinfo {author} {\bibfnamefont {H.}~\bibnamefont {Menke}}, \bibinfo {author} {\bibfnamefont {A.}~\bibnamefont {Urvoy}}, \bibinfo {author} {\bibfnamefont {O.}~\bibnamefont {Firstenberg}}, \bibinfo {author} {\bibfnamefont {H.~P.}\ \bibnamefont {B{\"u}chler}},\ and\ \bibinfo {author} {\bibfnamefont {S.}~\bibnamefont {Hofferberth}},\ }\bibfield  {title} {\bibinfo {title} {{Tutorial: Calculation of Rydberg interaction potentials}},\ }\href {https://doi.org/10.1088/1361-6455/aa743a} {\bibfield  {journal} {\bibinfo  {journal} {J. Phys. B: At. Mol. Opt. Phys.}\ }\textbf {\bibinfo {volume} {50}},\ \bibinfo {pages} {133001} (\bibinfo {year} {2017})}\BibitemShut {NoStop}%
\bibitem [{\citenamefont {Marder}\ \emph {et~al.}(1990)\citenamefont {Marder}, \citenamefont {Papanicolaou},\ and\ \citenamefont {Psaltakis}}]{Marder1990}%
  \BibitemOpen
  \bibfield  {author} {\bibinfo {author} {\bibfnamefont {M.}~\bibnamefont {Marder}}, \bibinfo {author} {\bibfnamefont {N.}~\bibnamefont {Papanicolaou}},\ and\ \bibinfo {author} {\bibfnamefont {G.~C.}\ \bibnamefont {Psaltakis}},\ }\bibfield  {title} {\bibinfo {title} {{Phase separation in a t-J model}},\ }\href {https://doi.org/10.1103/physrevb.41.6920} {\bibfield  {journal} {\bibinfo  {journal} {Physical Review B}\ }\textbf {\bibinfo {volume} {41}},\ \bibinfo {pages} {6920} (\bibinfo {year} {1990})}\BibitemShut {NoStop}%
\bibitem [{\citenamefont {Bobroff}\ \emph {et~al.}(2002)\citenamefont {Bobroff}, \citenamefont {Alloul}, \citenamefont {Ouazi}, \citenamefont {Mendels}, \citenamefont {Mahajan}, \citenamefont {Blanchard}, \citenamefont {Collin}, \citenamefont {Guillen},\ and\ \citenamefont {Marucco}}]{Bobroff2002}%
  \BibitemOpen
  \bibfield  {author} {\bibinfo {author} {\bibfnamefont {J.}~\bibnamefont {Bobroff}}, \bibinfo {author} {\bibfnamefont {H.}~\bibnamefont {Alloul}}, \bibinfo {author} {\bibfnamefont {S.}~\bibnamefont {Ouazi}}, \bibinfo {author} {\bibfnamefont {P.}~\bibnamefont {Mendels}}, \bibinfo {author} {\bibfnamefont {A.}~\bibnamefont {Mahajan}}, \bibinfo {author} {\bibfnamefont {N.}~\bibnamefont {Blanchard}}, \bibinfo {author} {\bibfnamefont {G.}~\bibnamefont {Collin}}, \bibinfo {author} {\bibfnamefont {V.}~\bibnamefont {Guillen}},\ and\ \bibinfo {author} {\bibfnamefont {J.-F.}\ \bibnamefont {Marucco}},\ }\bibfield  {title} {\bibinfo {title} {{Absence of Static Phase Separation in the High ${T}_{c}$ Cuprate ${\mathrm{Y}\mathrm{B}\mathrm{a}}_{2}{\mathrm{C}\mathrm{u}}_{3}{\mathrm{O}}_{6+y}$}},\ }\href {https://doi.org/10.1103/physrevlett.89.157002} {\bibfield  {journal} {\bibinfo  {journal} {Physical Review Letters}\ }\textbf {\bibinfo {volume} {89}},\ \bibinfo {pages} {157002} (\bibinfo {year} {2002})}\BibitemShut
  {NoStop}%
\bibitem [{\citenamefont {Daley}(2014)}]{Daley2014}%
  \BibitemOpen
  \bibfield  {author} {\bibinfo {author} {\bibfnamefont {A.~J.}\ \bibnamefont {Daley}},\ }\bibfield  {title} {\bibinfo {title} {Quantum trajectories and open many-body quantum systems},\ }\href {https://doi.org/10.1080/00018732.2014.933502} {\bibfield  {journal} {\bibinfo  {journal} {Advances in Physics}\ }\textbf {\bibinfo {volume} {63}},\ \bibinfo {pages} {77} (\bibinfo {year} {2014})}\BibitemShut {NoStop}%
\bibitem [{\citenamefont {Hauschild}\ \emph {et~al.}(2018)\citenamefont {Hauschild}, \citenamefont {Mong}, \citenamefont {Pollmann}, \citenamefont {Schulz}, \citenamefont {Schoonderwoert}, \citenamefont {Unfried}, \citenamefont {Tzeng},\ and\ \citenamefont {Zaletel}}]{Hauschild2018}%
  \BibitemOpen
  \bibfield  {author} {\bibinfo {author} {\bibfnamefont {J.}~\bibnamefont {Hauschild}}, \bibinfo {author} {\bibfnamefont {R.}~\bibnamefont {Mong}}, \bibinfo {author} {\bibfnamefont {F.}~\bibnamefont {Pollmann}}, \bibinfo {author} {\bibfnamefont {M.}~\bibnamefont {Schulz}}, \bibinfo {author} {\bibfnamefont {L.}~\bibnamefont {Schoonderwoert}}, \bibinfo {author} {\bibfnamefont {J.}~\bibnamefont {Unfried}}, \bibinfo {author} {\bibfnamefont {Y.}~\bibnamefont {Tzeng}},\ and\ \bibinfo {author} {\bibfnamefont {M.}~\bibnamefont {Zaletel}},\ }\bibfield  {title} {\bibinfo {title} {Tensor network python},\ }\href@noop {} {\bibfield  {journal} {\bibinfo  {journal} {\normalfont The code is available online at https://github.com/tenpy/tenpy/, the documentation can be found at https://tenpy.github.com/.}\ } (\bibinfo {year} {2018})}\BibitemShut {NoStop}%
\bibitem [{\citenamefont {Hauschild}\ and\ \citenamefont {Pollmann}(2018)}]{Hauschild2019}%
  \BibitemOpen
  \bibfield  {author} {\bibinfo {author} {\bibfnamefont {J.}~\bibnamefont {Hauschild}}\ and\ \bibinfo {author} {\bibfnamefont {F.}~\bibnamefont {Pollmann}},\ }\bibfield  {title} {\bibinfo {title} {Efficient numerical simulations with {T}ensor {N}etworks: {T}ensor {N}etwork {P}ython ({TeNPy})},\ }\bibfield  {journal} {\bibinfo  {journal} {{SciPost} Physics Lecture Notes}\ }\href {https://doi.org/10.21468/scipostphyslectnotes.5} {10.21468/scipostphyslectnotes.5} (\bibinfo {year} {2018})\BibitemShut {NoStop}%
\end{thebibliography}
\end{document}